\documentclass[twocolumn,prb,superscriptaddress,showkeys,preprintnumbers]{revtex4}
\usepackage{amssymb,graphicx}
\begin{document}

\title{Elastic theory of low-dimensional continua and its applications in bio- and nano-structures}
\author{Z. C. Tu}\email{tuzc@bnu.edu.cn}
\affiliation{Department of Physics, Beijing Normal University,
Beijing 100875, China} \affiliation{\textrm{II.} Institut f\"{u}r
Theoretische Physik, Universit\"{a}t Stuttgart, Pfaffenwaldring 57,
70550 Stuttgart, Germany}
\author{Z. C. Ou-Yang}\email{oy@itp.ac.cn}
\affiliation{Institute of Theoretical Physics, Chinese Academy of
Sciences, Beijing 100080, China}

\begin{abstract}
This review presents the elastic theory of low-dimensional (one- and
two-dimensional) continua and its applications in bio- and
nano-structures.

First, the curve and surface theory, as the geometric representation
of the low-dimensional continua, is briefly described through Cartan
moving frame method. The elastic theory of Kirchhoff rod, Helfrich
rod, bending-soften rod, fluid membrane, and solid shell is
revisited. The free energy density of the continua, is constructed
on the basis of the symmetry argument. The fundamental equations can
be derived from two kinds of viewpoints: the bottom-up and the
top-down standpoints. In the former case, the force and moment
balance equations are obtained from Newton's laws and then some
constitute relations are complemented in terms of the free energy
density. In the latter case, the fundamental equations are derived
directly from the variation of the free energy. Although the
fundamental equations have different forms obtained from these two
viewpoints, several examples reveal that they are, in fact,
equivalent to each other.

Secondly, the application and availability of the elastic theory of
low-dimensional continua in bio-structures, including short DNA
rings, lipid membranes, and cell membranes, are discussed. The kink
stability of short DNA rings is addressed by using the theory of
Kirchhoff rod, Helfrich rod, and bending-soften rod. The lipid
membranes obey the theory of fluid membrane. The shape equation and
the stability of closed lipid vesicles, the shape equation and
boundary conditions of open lipid vesicles with free edges as well
as vesicles with lipid domains, and the adhesions between a vesicle
and a substrate or another vesicle are fully investigated. A cell
membrane is simplified as a composite shell of lipid bilayer and
membrane skeleton, which is a little similar to the solid shell. The
equations to describe the in-plane strains and shapes of cell
membranes are obtained. It is found that the membrane skeleton
enhances highly the mechanical stability of cell membranes.

Thirdly, the application and availability of the elastic theory of
low-dimensional continua in nano-structures, including graphene and
carbon nanotubes, are discussed. A revised Lenosky lattice model is
proposed based on the local density approximation. Its continuum
form up to the second order terms of curvatures and strains is the
same as the free energy of 2D solid shells. The intrinsic roughening
of graphene and several typical mechanical properties of carbon
nanotubes are revisited and investigated based on this continuum
form. It is possible to avoid introducing the controversial
concepts, the Young's modulus and thickness of graphene and
single-walled carbon nanotubes, with this continuum form.


\preprint{J. Comput. Theor. Nanosci. 5, 422-448 (2008) \hspace{3cm}}
\end{abstract}\startpage{422}\maketitle


\section{Introduction}
We human beings live in a three-dimensional (3D) space which
contains many geometric entities composed of atoms or molecules. The
length scale of objects observed with our naked eyes is much larger
than the distance between nearest neighbor atoms or molecules in the
objects. As a result, the objects can be regarded as continua. If
one dimension of an object is much larger than the other two
dimensions, such as a rod, we call it a one-dimensional (1D) entity.
If one dimension of an object is much smaller than the other two
dimensions, such as a thin film, we call it a two-dimensional (2D)
entity. In this review, the term ``low-dimensional continua"
represents 1D and 2D entities.

Elasticity is a property of materials. It means that materials
deform under external forces, but return to their original shapes
when the forces are removed. Elastic theory, the study on the
elasticity of continuum materials, has a long history
\cite{Godoy,Love44} which records many geniuses such as Hooke
(1635--1703), Bernoulli (1700--1782), Euler (1707--1783), Lagrange
(1736--1813), Young (1773--1829), Poisson (1781--1840), Navier
(1785--1836), Cauchy (1789-1857), Green (1793--1841), Lam\'{e}
(1795--1870), Saint-Venant (1797--1886), Stokes (1819--1903),
Kirchhoff (1824--1887), and so on. Now elastic theory has been a
mature branch of physics and summarized in several excellent
textbooks. \cite{Love44,landau,Timoshenko84} Although the classical
elastic theory is applied to macroscopic continuum materials, more
and more facts reveal that it can be also available for bio- or
nano-structures such as short DNA rings,
\cite{TanakaJCP85,Zhaow98,ZhouPRE98,ZhouJCP99,Fain99,PanyukovPRE2001,ZhangSPRE04,ZhaoSPRE06,FainPRE97}
$\alpha$-helical coiled coils, \cite{SunPRL06} chiral filaments,
\cite{SmithPRL01,Kessler03,ZhouMPL05,ZhouZPRE05,WadaNetz07,LiuPLA03,LiuPLA06}
climbing plants, \cite{Goriely98,GorielyPRL06} bacterial flagella,
\cite{Goldstein2000} viral shells,
\cite{LidmarPRE03,NguyenPRE05,KlugPRL06} bio-membranes,
\cite{Canham,Helfrich73,Evans73,JenkinsJAM77,Lipowsky91,Seifert97,oybook,Zhong-canTSF,tzcAAPPS}
zinc oxide nanoribbons, \cite{Kongxy,Hugheswl,TuLiHu} and carbon
nanotubes,
\cite{Yakobson,Lujp,OuYangPRL97,PopovPRB2000,TuzcPRB02,RafiiTabarPR04,QianAMR02}
to some extent.

This review presents the elastic theory of low-dimensional continua
and its applications in bio- and nano-structures, which is organized
as follows: In Sec.~\ref{ETLDC}, we briefly introduce the geometric
representation and the elastic theory of low-dimensional continua
including 1D rod and 2D fluid membrane or solid shell. The free
energy density of the continua is constructed on the basis of the
symmetry argument. The fundamental equations can be derived from the
bottom-up and the top-down viewpoints. Although they have different
forms obtained from these two standpoints, several examples reveal
that they are, in fact, equivalent to each other. In
Sec.~\ref{SecApBioStuct}, the application and availability of the
elastic theory of low-dimensional continua in bio-structures,
including short DNA rings, lipid membranes, and cell membranes, are
discussed. We investigate the kink stability of short DNA rings, the
elasticity of lipid membranes, and the adhesions between a vesicle
and a substrate or another vesicle. A cell membrane is simplified as
a composite shell of lipid bilayer and membrane skeleton. The
membrane skeleton is shown to enhance highly the mechanical
stability of cell membranes. In Sec.~\ref{SecApNanoStuct}, the
application and availability of the elastic theory of
low-dimensional continua in nano-structures, including graphene and
carbon nanotubes, are discussed. We propose a revised Lenosky
lattice model and fit four parameters in this model through the
local density approximation. We derive its continuum form up to the
second order terms of curvatures and strains, which is the same as
the free energy of 2D solid shells. The intrinsic roughening of
graphene and several typical mechanical properties of carbon
nanotubes are revisited and investigated by using this continuum
form. Sec.~\ref{SecConclusion} is a brief summary and prospect.

\section{Fundamentals of geometric and elastic theory on low-dimensional continua \label{ETLDC}}
In this section, we describe the mathematical basis and the elastic
theory of 1D and 2D continua.
\subsection{Geometric representation of low-dimensional continua}
The 1D continuum (rod) and 2D continuum (membrane or shell) can be
expressed as a smooth curve and a smooth surface, respectively.
\subsubsection{Curve theory\label{curvesec}}
\begin{figure}[pth!]
\includegraphics[width=5cm]{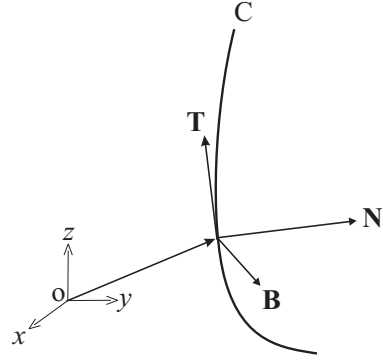}\caption{\label{figfrenet}
Frenet frame $\{\mathbf{r};\mathbf{T},\mathbf{N},\mathbf{B}\}$.}
\end{figure}

Fig.~\ref{figfrenet} depicts a curve $C$ embedded in the 3D Euclid
space. Each point in the curve can be expressed as a vector
$\mathbf{r}$ and let $s$ be the arc length parameter. At point
$\mathbf{r}(s)$, one can take $\mathbf{T}$, $\mathbf{N}$, and
$\mathbf{B}$ as the tangent, normal and binormal vectors,
respectively. $\{\mathbf{r};\mathbf{T},\mathbf{N},\mathbf{B}\}$ is
called the Frenet frame which satisfies the Frenet
formula:\cite{Carmobook}
\begin{equation}\left\{\begin{array}{l}\mathbf{r}'=\mathbf{T},\\
\mathbf{T}'=\kappa\mathbf{N},\\
\mathbf{N}'=(-\kappa\mathbf{T}+\tau\mathbf{B}),\\
\mathbf{B}'=-\tau\mathbf{N},\end{array}\right.\label{framefrenet}
\end{equation}
where the prime represents the derivative with respect to $s$.
$\kappa$ and $\tau$ are the curvature and torsion of the curve,
respectively.

The fundamental theory of curve \cite{Carmobook} tells us that the
bending and twist properties of a smooth curve are uniquely
determined by the Frenet formula (\ref{framefrenet}).

\subsubsection{Surface theory}
\begin{figure}[pth!]
\includegraphics[width=5cm]{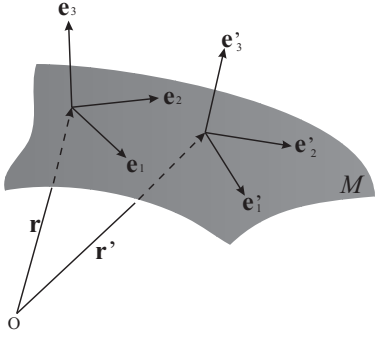}\caption{\label{sframefig}
Moving frame $\{\mathbf{r};\mathbf{e}_1,\mathbf{e}_2,\mathbf{e}_3\}$
of a surface $M$.}
\end{figure}

Fig.~\ref{sframefig} depicts a surface $M$ embedded in the 3D Euclid
space. Imagine that a mass point moves on the surface in the speed
of unit and that a right-handed frame, which consists of three unit
orthonormal vectors with two vectors always in the tangent plane of
the surface, adheres to the mass point. Assume that the mass point
is at position expressed as vector $\mathbf{r}$ and the frame
superposes three unit orthonormal vectors
$\{\mathbf{e}_1,\mathbf{e}_2,\mathbf{e}_3\}$ with $\mathbf{e}_3$
being the normal vector of surface $M$ at some time $s$. When the
mass point moves to another position $\mathbf{r}'$ at time $s+\Delta
s$, the frame will superpose three unit orthonormal vectors
$\{\mathbf{e}_1',\mathbf{e}_2',\mathbf{e}_3'\}$. Thus we call the
frame a moving frame and denote it as
$\{\mathbf{r};\mathbf{e}_1,\mathbf{e}_2,\mathbf{e}_3\}$.

If $\Delta s\rightarrow 0$, we define
\begin{equation}
d\mathbf{r}=\lim_{\Delta s\rightarrow
0}(\mathbf{r}'-\mathbf{r})=\omega_1\mathbf{e}_1+\omega_2\mathbf{e}_2,\label{sframer}
\end{equation}
and
\begin{equation}d\mathbf{e}_i=\lim_{\Delta s\rightarrow
0}(\mathbf{e}_i'-\mathbf{e}_i)=\omega_{ij}\mathbf{e}_j,\quad
(i=1,2,3)\label{sframee}\end{equation} where $\omega_1$, $\omega_2$,
and $\omega_{ij},(i,j=1,2,3)$ are 1-forms, and `$d$' is the exterior
differential operator. \cite{Chernbook,TuJPA04} Here $\omega_{12}$
can be understood as the infinite rotation angle of vectors
$\mathbf{e}_1$ and $\mathbf{e}_2$ around $\mathbf{e}_3$. Similarly,
we can understand the physical meaning of the other $\omega_{ij}$.
It is easy to obtain $\omega_{ij}=-\omega_{ji}$ from
$\mathbf{e}_i\cdot\mathbf{e}_j=\delta_{ij}$. Additionally, the
structure equations of the surface can be expressed as:
\cite{Chernbook,TuJPA04}
\begin{equation}
\left\{\begin{array}{l}d\omega_1=\omega_{12}\wedge\omega_2,\\
d\omega_2=\omega_{21}\wedge\omega_1,\\
\label{domgaij} d\omega_{ij}=\omega_{ik}\wedge\omega_{kj}\quad
(i,j=1,2,3),\end{array}\right. \label{structur}\end{equation} and
\begin{equation}
\left(\begin{array}{l}\omega_{13}\\
\omega_{23}\end{array}\right)=\left(\begin{array}{cc}a&b\\
b&c\end{array}\right)\left(\begin{array}{l}\omega_{1}\\
\omega_{2}\end{array}\right),\label{omega13}\end{equation} where
`$\wedge$' represents the wedge production between two differential
forms. The matrix $\left(\begin{array}{cc}a&b\\
b&c\end{array}\right)$ is the representation matrix of the curvature
tensor $\mathfrak{R}$. Its trace and determinant are two invariants
under the coordinate rotation around $\mathbf{e}_3$ which are
denoted by
\begin{equation}2H=a+c\quad \mathrm{and}\quad K=ac-b^2.\label{relat2HK}\end{equation}
They can be expressed as $2H=-(1/R_1+1/R_2)$ and $K=1/R_1R_2$ by the
two principal curvature radii $R_1$ and $R_2$ at each point.

Consider a tangent vector $\mathbf{m}$ stemming from $\mathbf{r}$.
Let $\phi$ be the angle between $\mathbf{m}$ and $\mathbf{e}_1$.
Then the geodesic curvature, the geodesic torsion, and the normal
curvature along the direction of $\mathbf{m}$ can be expressed:
\cite{TuJPA04}
\begin{equation}
\left\{\begin{array}{l}
k_{g}=(d\phi +\omega _{12})/ds,\\
\tau _{g} =b\cos2\phi+(c-a)\cos\phi\sin\phi,\\
k_{n}=a\cos^2\phi+2b\cos\phi\sin\phi+c\sin^2\phi,\end{array}\right.\label{geodisicc}
\end{equation}
where $ds$ is the arc length element along $\mathbf{m}$. If
$\mathbf{m}$ aligns with $\mathbf{e}_1$, then $\phi=0$,
$k_{g}=\omega _{12}/ds$, $\tau _{g} =b$, and $k_{n}=a$.

\begin{figure}[pth!]
\includegraphics[width=5cm]{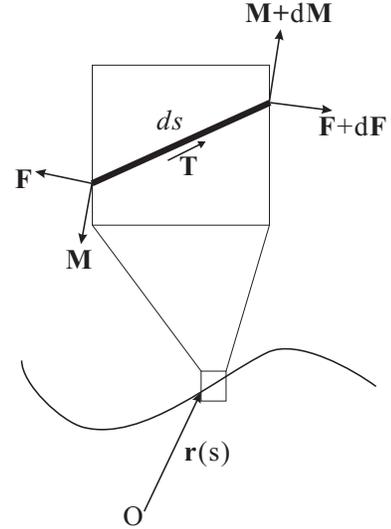}\caption{\label{figKrod-elm}
Force and moment in 1D rod.}
\end{figure}

\subsection{Elastic theory of 1D continua \label{Sectionrod}}
We will elucidate the elastic theory of rod with inextensible
centerline. As shown in Fig.~\ref{figKrod-elm}, let us simplify a
rod as a curve $\mathbf{r}(s)$ with $s$ being the arc-length
parameter, and cut an infinitesimal element (shown in the magnified
box) from the rod. There are forces and moments at the two ends of
the element which originating from the interaction of other parts of
the rod. $\mathbf{F}$ and $\mathbf{M}$ represent the force and
moment vectors at point $\mathbf{r}(s)$, while
$\mathbf{F}+d\mathbf{F}$ and $\mathbf{M}+d\mathbf{M}$ are the force
and moment vectors at point $\mathbf{r}(s+ds)$. From Newton's laws,
we can derive the force and moment balance equations:
\begin{equation}\sum \mathbf{F}=0\Rightarrow \mathbf{F}'=0,\label{forcebal}\end{equation}
and
\begin{equation}\sum \mathbf{M}=0\Rightarrow \mathbf{M}'+\mathbf{T}\times\mathbf{F}=0,\label{momentbal}\end{equation}
where the prime represents the derivative with respect to $s$. One
should add the constitutive relation and boundary conditions to make
the above two equations closed.

\subsubsection{Kirchhoff rod theory}
\begin{figure}[pth!]
\includegraphics[width=7cm]{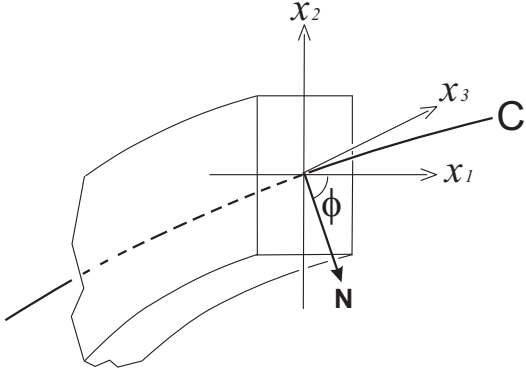}\caption{\label{figRodsec}
Rod with rectangle cross section.}
\end{figure}

A rod with rectangle cross section and centerline $C$ is shown in
Fig.~\ref{figRodsec}. Take local coordinates $\{x_1,x_2,x_3\}$ with
$x_1$ and $x_2$ paralleling respectively to the two edges of the
rectangle, and $x_3$ along the tangent of the centerline.
$\mathbf{N}$ is the normal of curve $C$. Let
$\{\mathbf{x}_1,\mathbf{x}_2,\mathbf{x}_3\}$ denote the basis of the
local coordinates and define
$\kappa_1=-\mathbf{x}_2\cdot(d\mathbf{x}_3/ds)$,
$\kappa_2=\mathbf{x}_1\cdot(d\mathbf{x}_3/ds)$, and
$\kappa_3=\mathbf{x}_2\cdot(d\mathbf{x}_1/ds)$. Viewed from
geometrical point, $\kappa_1$ and $\kappa_2$ describe the bending of
the rod around axes $x_1$ and $x_3$, respectively, and $\kappa_3$
represents the twist of the rod around axis $x_3$. The free energy
density $G$ due to the bending and twist can be expressed as a
function of $\kappa_1$, $\kappa_2$, and $\kappa_3$. Expanding $G$ up
to the second order terms of $\kappa_1$, $\kappa_2$, and $\kappa_3$,
we have
\begin{equation}G=\gamma+\frac{k_1}{2}(\kappa_1-\bar{\kappa}_1)^2+\frac{k_2}{2}(\kappa_2-\bar{\kappa}_2)^2+\frac{k_2}{2}(\kappa_3-\bar{\kappa}_3)^2,\label{K-rod1}\end{equation}
where the constant $\gamma$ can be interpreted as the line tension.
$\bar{\kappa}_1$ and $\bar{\kappa}_2$ are interpreted as the
spontaneous curvatures while $\bar{\kappa}_3$ the spontaneous
torsion. Denote
$\mathbf{k}=\kappa_1\mathbf{x}_1+\kappa_2\mathbf{x}_2+\kappa_3\mathbf{x}_3$
and let $\phi$ be the angle between $\mathbf{x}_1$ and $\mathbf{N}$.
Then we have
\begin{equation}\left\{\begin{array}{l}\mathbf{N}=\cos\phi\, \mathbf{x}_1-\sin\phi\, \mathbf{x}_2,\\
\mathbf{B}=\sin\phi\, \mathbf{x}_1+\cos\phi\,
\mathbf{x}_2,\end{array}\right.\label{tranfcood}\end{equation} where
$\mathbf{B}$ is the binormal of curve $C$. From
Eqs.~(\ref{framefrenet}) and (\ref{tranfcood}), we can derive
\cite{McMillenJNS02,ShipmanPRE02}
\begin{equation}\mathbf{k}=\kappa\sin\phi\,\mathbf{x}_1+\kappa
\cos\phi\,\mathbf{x}_2+(\tau+\phi')\,\mathbf{x}_3.\label{kvexprs}\end{equation}
Thus $G$ can be also regarded as the function of
$\kappa,\tau,\phi,\phi'$.

The moment vector is defined as \cite{Love44}
\begin{equation}\mathbf{M}=\frac{\partial G}{\partial \mathbf{k}}\equiv \frac{\partial G}{\partial\kappa_1}\,\mathbf{x}_1+\frac{\partial G}{\partial\kappa_2}\,\mathbf{x}_2+\frac{\partial G}{\partial\kappa_3}\,\mathbf{x}_3,\label{momentvec}\end{equation}
which is called the constitutive relation. Eqs.~(\ref{forcebal}),
(\ref{momentbal}) and (\ref{momentvec}) with some boundary
conditions form a group of closed equations. They are also available
for the rod with cross section different from rectangle if only we
take $x_1$ and $x_2$ as the two principal axes of inertia. It should
be noted that the equivalent form of these equations can be also
obtained from the variational method. This method is called the
top-down method while the former one via Newton's laws called the
bottom-up method.

The free energy of a rod with length $L$ can be written as
\begin{equation}\mathcal{F}=\int_0^L G(\kappa,\tau;\phi,\phi')\,ds+\mathcal{F}_{bd},\label{freeng-rod}\end{equation}
where $\mathcal{F}_{bd}$ comes from the contributions of two ends of
the rod. The general Euler-Lagrange equations corresponding to
Eq.~(\ref{freeng-rod}) are derived as
\begin{eqnarray}
&&\hspace{-0.9cm}G_{\phi }-( G_{\phi ^{\prime }}) ^{\prime } =0,\label{ELKrod1}\\
&&\hspace{-0.9cm}G_{\kappa }^{\prime \prime }+2\tau ( G_{\tau
}^{\prime }/\kappa) ^{\prime }+G_{\tau }^{\prime }\tau ^{\prime
}/\kappa +( \kappa
^{2}-\tau ^{2}) G_{\kappa }\nonumber\\
&&\hspace{-0.35cm}+2\kappa \tau G_{\tau}+\kappa \phi
^{\prime }G_{\phi ^{\prime }}-\kappa G =0,\label{ELKrod2}\\
&&\hspace{-0.9cm}\tau ^{\prime }G_{\kappa }+2\tau G_{\kappa
}^{\prime }-( \kappa G_{\tau }) ^{\prime }+( \tau ^{2}/\kappa)
G_{\tau }^{\prime }-( G_{\tau }^{\prime }/\kappa) ^{\prime \prime }
=0,\label{ELKrod3}
\end{eqnarray}
where $G_\phi$, $G_{\phi'}$, $G_\kappa$ and $G_\tau$ are the partial
derivatives of $G$ with respect to $\phi$, ${\phi'}$, $\kappa$ and
$\tau$, respectively. Additionally,
$G_{\kappa}'\equiv(G_{\kappa})'$, $G_{\tau}'\equiv(G_{\tau})'$,
$G_{\kappa}''\equiv(G_{\kappa})''$. The berief derivation of
Eqs.~(\ref{ELKrod1})--(\ref{ELKrod3}) is attached in Appendix
\ref{SecAppvarcc}. These equations have been employed to investigate
helical and twisted filaments. \cite{ZhaoSPRE06} There might be a
misprint in Eq.~(7) of Ref.~\onlinecite{ZhaoSPRE06}, corresponding
to our above equation (\ref{ELKrod3}), because the dimension of its
last term is different from that of other terms.

Now we would give a typical example to reveal the equivalence
relation between Eqs.~(\ref{forcebal}),(\ref{momentbal}),
(\ref{momentvec}) and Eqs.~(\ref{ELKrod1})--(\ref{ELKrod3}) rather
than prove it directly. Let us consider a rod with $k_1=k_2=k_{0}$,
$k_3=0$, and $\bar{\kappa}_1=\bar{\kappa}_2=\bar{\kappa}_3=0$. The
free energy density (\ref{K-rod1}) is simplified as
\begin{equation}G =( k_{0}/2) (\kappa _{1}^{2}+\kappa _{1}^{2})+\gamma =(
k_{0}/2) \kappa ^{2}+\gamma.\label{purebend0}\end{equation} On the
one hand, we have $M_{1}
=k_{0}\kappa_{1}=k_{0}\kappa\sin\phi,M_{2}=k_{0}\kappa
_{2}=k_{0}\kappa\cos\phi,M_{3}=0$ from Eq.~(\ref{momentvec}). The
moment balance equation (\ref{momentbal}) implies
$F_{1}=-k_{0}\kappa _{1}\kappa _{3}-k_{0}\kappa _{2}^{\prime }$ and
$F_{2}=k_{0}\kappa _{1}^{\prime }-k_{0}\kappa _{2}\kappa _{3}$.
Substituting them into the force balance equation (\ref{forcebal}),
we have $F_{3}=F_{30}-k_{0}\kappa ^{2}/2$ and
\begin{eqnarray}&&\kappa ^{\prime \prime }-\kappa \tau
^{2}+\kappa ^{3}/2-\kappa F_{30}/k_{0}=0,\label{string1}\\
&&2\tau \kappa ^{\prime }+\kappa \tau ^{\prime
}=0,\label{string2}\end{eqnarray} where $F_{30}$ is an integral
constant which represents the line tension of the straight
($\kappa=0$) rod. On the other hand, we have $G_{\kappa
}=k_{0}\kappa$, $G_{\phi}=G_{\phi ^{\prime }}=G_{\tau }=0$.
Eq.~(\ref{ELKrod1}) is trivial while Eqs.~(\ref{ELKrod2}) and
(\ref{ELKrod3}) are, respectively, transformed into
\begin{eqnarray}
&&\kappa ^{\prime \prime }-\kappa \tau ^{2}+\kappa ^{3}/2-\gamma
\kappa/k_{0}=0,\label{string11}\\
&&2\tau \kappa ^{\prime }+\kappa \tau ^{\prime }=0.\label{string22}
\end{eqnarray}
The above equations are the same as Eqs.~(\ref{string1}) and
(\ref{string2}) obtained from the force and moment balance
conditions if only we take $F_{30}=\gamma$. Thus the equations
obtained from the top-down and bottom-up methods are equivalent to
each other.

Substituting the free energy density (\ref{K-rod1}) into
Eqs.~(\ref{ELKrod1})--(\ref{ELKrod3}), we obtain the so called shape
equations of Kirchhoff rod as
\begin{eqnarray}&&( k_{1}-k_{2}) \kappa ^{2}\sin 2\phi
-2k_{3}(\tau +\phi ^{\prime }) ^{\prime }+2I_{21}
\kappa=0,\label{RodShape1}\\ \nonumber\\ &&\quad I_1 ( 2\kappa
^{\prime \prime }+\kappa ^{3}-2\kappa \tau ^{2})
-2\gamma \kappa+2I_{12} ( \phi ^{\prime 2}+\tau ^{2})  \nonumber\\
&&+2I_{21} \phi ^{\prime \prime }+2( k_{1}-k_{2}) [ ( \phi ^{\prime
}\kappa \sin 2\phi ) ^{\prime }+\phi ^{\prime }\kappa ^{\prime }\sin
2\phi ] \nonumber\\
&& -\bar{I} \kappa+4k_{3}\tau [ ( \tau ^{\prime }+\phi ^{\prime
\prime }) /\kappa ] ^{\prime }+2k_{3}( \tau ^{\prime }+\phi ^{\prime
\prime }) \tau
^{\prime }/\kappa\nonumber\\
&&+k_{3}\kappa ( \tau +\phi ^{\prime }-\bar{\kappa}_{3}) ( 3\tau
+\phi ^{\prime }+\bar{\kappa}_{3})=0,\label{RodShape2}
\\ \nonumber \\
&&\quad I_{1}(\tau ^{\prime }\kappa +2\tau \kappa ^{\prime }) -k_{3}
[\kappa ( \tau +\phi ^{\prime }-\bar{\kappa}_{3})]
^{\prime }\nonumber \\
&&+k_{3}\tau ^{2}( \tau ^{\prime }+\phi ^{\prime \prime }) /\kappa
-k_{3}[( \tau ^{\prime }+\phi ^{\prime \prime })
/\kappa] ^{\prime \prime }\nonumber \\
&&+2\tau [( k_{1}-k_{2}) \kappa \sin 2\phi +I_{21}] \phi ^{\prime
}-I_{12}\tau ^{\prime }=0,\label{RodShape3}
\end{eqnarray}
where $I_1=k_{1}\sin ^{2}\phi +k_{2}\cos ^{2}\phi$,
$\bar{I}=k_{1}\bar{\kappa}_1^2+k_{2}\bar{\kappa}_2^2$,
$I_{12}=k_{1}\bar{\kappa}_1\sin\phi+k_{2}\bar{\kappa}_2\cos\phi$,
and
$I_{21}=k_{2}\bar{\kappa}_2\sin\phi-k_{1}\bar{\kappa}_1\cos\phi$.

We also suggest that gentle readers consult the work by Zhou
\textit{et al.}\cite{ZhouZPRE05} where the above equations
(\ref{RodShape1})--(\ref{RodShape3}) and different kinds of boundary
conditions are expressed in another representation with the aid of
Euler angles.

\subsubsection{Helfrich rod theory}
Helfrich rod theory can be regarded as the fourth order Kirchhoff
rod theory with circular cross section to some extent. The free
energy density is expressed as \cite{HelfrichLangm90}
\begin{equation}
G=\frac{1}{2}k_{2}\kappa ^{2}+k_{3}\kappa ^{2}\tau
+\frac{1}{4}k_{22}\kappa ^{4}+\frac{1}{2}k_{4}({\kappa'}^{2}+\kappa
^{2}\tau ^{2})+\gamma,\label{FrenHlfRod}
\end{equation}
where $k_{2}$, $k_{3}$, $k_{22}$ and $k_{4}$ are elastic constants
while $\gamma$ is the line tension. It is noted that this free
energy density is the simplest stable form including the chirality
term but without spontaneous curvature and torsion. It has been
employed to investigate the circular DNA in
Ref.~\onlinecite{Zhaow98} and the Euler-Lagrange equations
corresponding to $\int G\, ds$ are given as:
\begin{eqnarray}
&&\quad k_{2}(\kappa ^{3}/2-\kappa \tau ^{2}+\kappa ^{\prime \prime
})-\gamma
\kappa  \nonumber\\
&&+k_{3}(3\kappa ^{3}\tau -2\kappa \tau ^{3}+6\kappa ^{\prime }\tau
^{\prime
}+2\kappa \tau ^{\prime \prime }+6\kappa ^{\prime \prime }\tau )  \nonumber\\
&&+k_{4}(5\kappa ^{3}\tau ^{2}/2-\kappa \tau ^{4}+\kappa \kappa
^{\prime 2}/2-\kappa ^{2}\kappa ^{\prime \prime }-\kappa ^{\prime
\prime \prime
\prime }  \nonumber\\
&&+6\kappa ^{\prime \prime }\tau ^{2}+12\kappa ^{\prime }\tau \tau
^{\prime
}+4\kappa \tau \tau ^{\prime \prime }+3\kappa \tau ^{\prime 2}) \nonumber \\
&&+k_{22}(3\kappa ^{5}/4-\kappa ^{3}\tau ^{2}+6\kappa \kappa
^{\prime 2}+3\kappa ^{2}\kappa ^{\prime \prime })=0,\label{HfRod1}\\
 \nonumber\\
&&\quad k_{2}(2\kappa ^{\prime }\tau +\kappa \tau ^{\prime
})+k_{22}(\kappa
^{3}\tau ^{\prime }+6\kappa ^{2}\kappa ^{\prime }\tau ) \nonumber \\
&&+k_{3}(6\kappa ^{\prime }\tau ^{2}+6\kappa \tau \tau ^{\prime
}-3\kappa
^{2}\kappa ^{\prime }-2\kappa ^{\prime \prime \prime })  \nonumber\\
&&+k_{4}(4\kappa ^{\prime }\tau ^{3}+6\kappa \tau ^{2}\tau ^{\prime
}-3\kappa ^{2}\kappa ^{\prime }\tau -\kappa ^{3}\tau ^{\prime }  \nonumber\\
&&-4\kappa ^{\prime }\tau ^{\prime \prime }-6\kappa ^{\prime \prime
}\tau ^{\prime }-4\kappa ^{\prime \prime \prime }\tau -\kappa \tau
^{\prime \prime \prime })=0.\label{HfRod2}
\end{eqnarray}

Here we will not go on the more higher order Helfrich rod theory, on
which gentle readers can consult Refs.~\onlinecite{LiuPLA03} and
\onlinecite{LiuPLA06}.

\subsubsection{Theory of bending-soften Rod}
\begin{figure}[pth!]
\includegraphics[width=7cm]{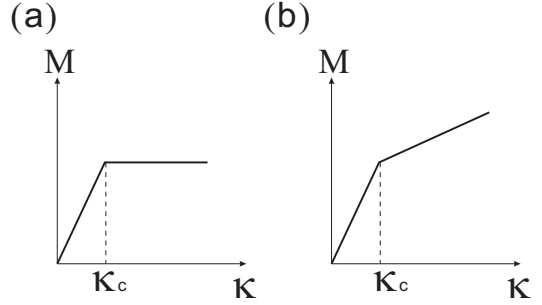}\caption{\label{figbendsoft}
Bending-induced softening relation: (a) the first kind in expression
of Eq.~(\ref{BSRodMom1}); (b) the second kind in expression of
Eq.~(\ref{BSRodMom2}).}
\end{figure}

There are two kinds of rod theory with bending-induced softening.
First, let us assume that the bending moment depends linearly on the
curvature for small curvature but not on the curvature for large
curvature, which is expressed as
\begin{equation}M=\left\{\begin{array}{l}k_1\kappa,\quad (\kappa<\kappa_c)\\ k_1\kappa_c,\quad (\kappa>\kappa_c)\end{array} \right.\label{BSRodMom1}\end{equation}
where $k_1$ and $\kappa_c$ are the elastic bending rigidity and the
critical curvature, respectively. Eq.~(\ref{BSRodMom1}) describes
the bending-induced softening relation of the first kind which is
depicted in Fig.~\ref{figbendsoft}(a). The corresponding free energy
density can be expressed as
\begin{equation}G=\gamma+(k_1/2)[\kappa^2-(\kappa-\kappa_c)^2\mathcal{H}(\kappa-\kappa_0)],\label{BSRodFEG1}\end{equation}
where $\mathcal{H}(.)$ is the Heaviside step function. The above
form has been employed by Yan \emph{et al.} to investigate the loop
formation mechanism and probability of short DNA
rings.\cite{YanMarko05} We conjecture that this model could solve
the paradox in the experiment on the ring closure of single-walled
carbon nanotubes with 1,3-dicyclohexylcarbodiimide.\cite{SanoSCI01}
Fitting the experiment data with the worm-like chain
model,\cite{YamakawaJCP72} the persistence length is 800~nm for
single-walled carbon nanotubes in the diameter of
1~nm,\cite{SanoSCI01} which is much smaller than the theoretical
value $33\, \mu$m estimated in terms of the Young's modulus and
thickness of single-walled carbon nanotubes in
Ref.~\onlinecite{TuzcPRB02}.

Consider a rod divided into two parts at $s=L_c$: one part ($s<L_c$)
has curvatures less than $\kappa_c$ another one larger than
$\kappa_c$. In terms of the variational method in Appendix
\ref{SecAppvarcc}, we can derive the equations describing the rod as
\begin{eqnarray}
&&k_{1}( 2\kappa ^{\prime \prime }-2\kappa \tau ^{2}+\kappa ^{3})
-2\gamma \kappa  =0\quad (s<L_c),\label{ELsoftrod1}\\
&&\kappa \tau ^{\prime }+2\kappa ^{\prime }\tau  =0\quad (s<L_c),\label{ELsoftrod2}\\
&&k_{1}\kappa _{c}( \kappa _{c}\kappa -2\tau ^{2}) -2\gamma \kappa
=0\quad (s>L_c), \label{ELsoftrod3}\\
&&\tau ^{\prime } =0\quad (s>L_c).\label{ELsoftrod4}
\end{eqnarray}
At the divided point $s=L_c$, we have the joint conditions as
\begin{eqnarray}
&&\kappa _{-}=\kappa _{+}=\kappa _{c}, \label{sftrdjtc1}\\
&&\kappa _{-}^{\prime } =0,  \label{sftrdjtc2}\\
&&\tau _{-} =\tau _{+}, \label{sftrdjtc3}
\end{eqnarray}
where $(.)_-$ and $(.)_+$ represent the values of $(.)$ at the left
and right sides of $s=L_c$.

Secondly, let us assume that the bending moment depends linearly on
the curvature for small curvature but weaker linearly on the
curvature for large curvature, which is expressed as
\begin{equation}M=\left\{\begin{array}{l}k_1\kappa,\quad (\kappa<\kappa_c)\\
k_2(\kappa-\kappa_c)+k_1\kappa_c,\quad (\kappa>\kappa_c)\end{array}
\right.\label{BSRodMom2}\end{equation} where $k_1>k_2$ are the
elastic bending rigidities while $\kappa_c$ is the critical
curvature. Eq.~(\ref{BSRodMom2}) describes the bending-induced
softening relation of the second kind which is depicted in
Fig.~\ref{figbendsoft}(b). The corresponding free energy density can
be expressed as
\begin{equation}G=\gamma+(k_1/2)\kappa^2+[(k_2-k_1)/2](\kappa-\kappa_c)^2\mathcal{H}(\kappa-\kappa_0),\label{BSRodFEG2}\end{equation}

Consider a rod divided into two parts at $s=L_c$: one part ($s<L_c$)
has curvatures less than $\kappa_c$ another one larger than
$\kappa_c$. In terms of the variational method in Appendix
\ref{SecAppvarcc}, we can derive the equations describing the rod as
\begin{eqnarray}
&&\hspace{-0.6cm}k_{1}( 2\kappa ^{\prime \prime }-2\kappa \tau
^{2}+\kappa ^{3})
-2\gamma \kappa =0\quad (s<L_c),\label{ELsoftrod11} \\
&&\hspace{-0.6cm}\kappa \tau ^{\prime }+2\kappa ^{\prime }\tau =0\quad (s<L_c),\label{ELsoftrod22} \\
&&\hspace{-0.6cm}2k_{2}\kappa ^{\prime \prime }+[ k_{2}( \kappa
-\kappa
_{c}) +k_{1}\kappa _{c}] ( \kappa ^{2}-2\tau ^{2})-2\gamma \kappa\nonumber  \\
&&\hspace{0.4cm} +( k_{2}-k_{1}) ( \kappa -\kappa _{c}) \kappa
\kappa
_{c}  =0\quad (s>L_{c}), \label{ELsoftrod33}\\
&&\hspace{-0.6cm}k_{2}( \tau ^{\prime }\kappa +2\kappa ^{\prime
}\tau ) +( k_{1}-k_{2}) \kappa _{c}\tau ^{\prime } =0\quad
(s>L_{c}).\label{ELsoftrod44}
\end{eqnarray}

At the divided point $s=L_c$, we have the joint conditions as
\begin{eqnarray}
&&k_{1}( \kappa _{-}-\kappa _{c})  =k_{2}( \kappa
_{+}-\kappa _{c}) \label{sftrdjtc11} \\
&&k_{1}\kappa _{-}^{\prime } =k_{2}\kappa _{+}^{\prime } \label{sftrdjtc22} \\
&&\tau _{-} =\tau _{+}  \label{sftrdjtc33}\\
&&k_{1}( \kappa _{-}^{2}-\kappa _{+}^{2}) =(k_{2}-k_{1})( \kappa
_{+}-\kappa _{c}) ^{2}. \label{sftrdjtc44}
\end{eqnarray}

Obviously, the above equations
(\ref{ELsoftrod11})--(\ref{sftrdjtc44}) degenerate into
Eqs.~(\ref{ELsoftrod1})--(\ref{sftrdjtc3}) if $k_2=0$ and into
Eqs.~(\ref{string11})--(\ref{string22}) if $k_2=k_1$.

\subsection{Elastic theory of 2D continua\label{Sec2Dcont}}
\begin{figure}[pth!]
\includegraphics[width=6cm]{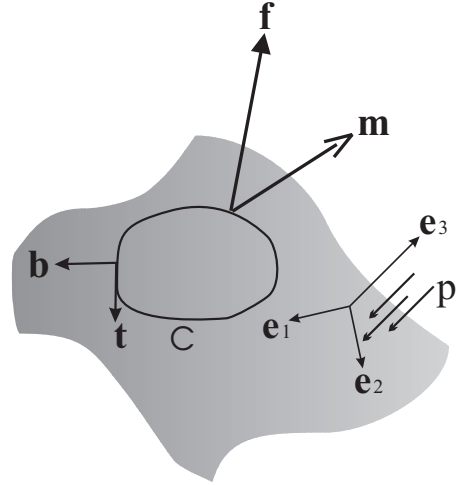}\caption{\label{figshellelm}
Force and moment in a 2D continuum.}
\end{figure}

A 2D continuum can be simplified as a surface as shown in
Fig.~\ref{figshellelm}. At each point, we can select a frame
$\{\mathbf{e}_1,\mathbf{e}_2,\mathbf{e}_3\}$. A pressure $p$ is
loaded on the surface in the inverse direction of the normal vector
$\mathbf{e}_3$. Let us cut a region enclosed in any curve $C$ from
the surface. $\mathbf{t}$ is the tangent vector at point of curve
$C$. $\mathbf{b}$ is normal to $\mathbf{t}$ and in the tangent
plane. The force and moment per length performed by the other region
on curve $C$ are denoted as $\mathbf{f}$ and $\mathbf{m}$,
respectively. Through Newton's laws, the force and moment balance
conditions are obtained as
\begin{eqnarray}
&&\oint_C \mathbf{f}\,ds-\int p\mathbf{e}_{3}\,dA =0,\label{forcb2D01} \\
&&\oint_C \mathbf{m}\,ds+\oint_C \mathbf{r}\times
\mathbf{f}\,ds-\int \mathbf{r} \times p\mathbf{e}_{3}\,dA
=0,\label{forcb2D02}
\end{eqnarray}
where $ds$ and $dA$ are the arc length element of curve $C$ and area
element of the region enclosed in curve $C$, respectively.

Define two second order tensors $\mathfrak{S}$ and $\mathfrak{M}$
such that
\begin{equation}\mathfrak{S}\cdot\mathbf{b}=\mathbf{f},\quad\mathfrak{M}\cdot\mathbf{b}=\mathbf{m}.\end{equation}
These two tensors can be called as stress tensor and bending moment
tensor, respectively. Using the Stokes' theorem, we can derive
\begin{eqnarray}&&\int (\mathrm{div\,}\mathfrak{S}-p\mathbf{e}_{3})\, dA=0,\\
&&\int
(\mathrm{div\,}\mathfrak{M}+\mathbf{e}_1\times\mathfrak{S}_1+\mathbf{e}_2\times\mathfrak{S}_2)\,dA=0.
\end{eqnarray}
where $\mathfrak{S}_1=\mathfrak{S}\cdot\mathbf{e}_1$ and
$\mathfrak{S}_2=\mathfrak{S}\cdot\mathbf{e}_2$. Since the integral
is performed on the region enclosed in an arbitrary curve $C$, from
the above two equations we obtain the force and moment balance
conditions of 2D continua as:
\begin{eqnarray}
&&\mathrm{div\,} \mathfrak{S} =p\mathbf{e}_{3},\label{forcb2D1} \\
&&\mathrm{div\,}\mathfrak{M} =\mathfrak{S}_{1}\times
\mathbf{e}_{1}+\mathfrak{S}_{2}\times \mathbf{e}_{2}.\label{momb2D1}
\end{eqnarray}
The above two equations are equivalent to Eq.~(25) in
Ref.~\onlinecite{ERICKSENTRU}, and Eqs.~(28) and (57) in
Ref.~\onlinecite{GuvenJPA02}. Eqs.~(\ref{forcb2D1}) and
(\ref{momb2D1}) with some complement constitutive relations form the
fundamental equations of 2D continua.
\subsubsection{Fluid membranes}
A fluid membrane is a 2D isotropic continuum which cannot withstand
in-plane shear strain. Generally, we assume that the fluid is
incompressible. The free energy density, $G$, of fluid membranes
should be invariant under the in-plane coordinate transformation. In
terms of the surface theory, there are only two fundamental
geometric invariants: the mean curvature $2H$ and gaussian curvature
$K$. Thus the free energy density should be a function of $2H$ and
$K$, that is,
\begin{equation}G=G(2H,K).\label{freeng-FM}\end{equation}

The free energy of a closed fluid membrane can be expressed as
\begin{equation}\mathcal{F}=\int G\, dA +p\int dV,\label{frengFM}\end{equation}
where $dA$ is the area element of the membrane and $dV$ is the
volume element enclosed in the membrane. $p$ is the osmotic
pressure, the pressure difference between the outer and inner side
of the membrane. The general Euler-Lagrange equation of free energy
(\ref{frengFM}) can be derived through the variational method shown
in Appendix \ref{varisurf} as
\begin{eqnarray}&&p-2HG+(\nabla^2/2+2H^2-K)(\partial G/\partial H)
\nonumber\\
&&\hspace{1.45cm}+(\nabla\cdot\tilde{\nabla}+2KH)(\partial
G/\partial K)=0.\label{ELclosFM}\end{eqnarray} As we known, the
above equation has been derived by several authors such as Ou-Yang
\textit{et al.} \cite{Naitopre95,TuJPA04} and Giaquinta \textit{et
al.} \cite{Giaquintabook96} coming from different research fields.
It is recently employed to investigate the modified Korteweg-de
Vries surfaces. \cite{TekJMP07} Here $\nabla\cdot\tilde{\nabla}$ can
be called as the Laplace operator of the second class which is also
fully discussed by Zhang and Xu.\cite{ZhangXu07}

We emphasize that (\ref{ELclosFM}) can be also derived from the
bottom-up method, Eqs.~(\ref{forcb2D1}) and (\ref{momb2D1})
combining a complement constitutive relation
\begin{equation}\mathfrak{M}=(G_{b}/2)( \mathbf{e}_{1}\mathbf{e}_{1}-
\mathbf{e}_{2}\mathbf{e}_{2}) -G_{a}\mathbf{e}_{2}\mathbf{e}_{1}+G_{c}%
\mathbf{e}_{1}\mathbf{e}_{2},\label{consrelation2}\end{equation}
where $G_a$, $G_b$, and $G_c$ represent the partial derivatives of
$G$ with respect to $a$, $b$, and $c$, respectively. Here $a$, $b$,
and $c$ are the components of the curvature tensor $\mathfrak{R}$ in
Eq.~(\ref{omega13}). To illuminate this point, we consider an
example in which the free energy density is taken as
$G=k_c(2H)^2+\lambda$, where $k_c$ and $\lambda$ are the bending
modulus and surface tension of the fluid membrane. It follows that
$\mathfrak{M}=2k_cH(\mathbf{e}_1\mathbf{e}_2-\mathbf{e}_2\mathbf{e}_1)$
from Eq.~(\ref{consrelation2}). Substituting it into
Eqs.~(\ref{forcb2D1}) and (\ref{momb2D1}), we can derive
\begin{equation}p-2\lambda H+4k_cH(H^{2}-K) +2k_c\nabla ^{2}H =0,\end{equation}
which is the same as the result obtained directly from
(\ref{ELclosFM}). Simultaneously, we have the stress components
\begin{eqnarray}&&\hspace{-0.4cm}\mathfrak{S}_1=(2H^2-2aH+\lambda)\mathbf{e}_1-2bH \mathbf{e}_2-2H_1\mathbf{e}_3,\label{stress1}\\
&&\hspace{-0.4cm}\mathfrak{S}_2=-2bH\mathbf{e}_1+(2H^2-2cH+\lambda)\mathbf{e}_2-2H_2\mathbf{e}_3,\label{stress2}
\end{eqnarray}
where $H_1$ and $H_2$ are the directional derivatives of $H$ respect
to $\mathbf{e}_1$ and $\mathbf{e}_2$. These equations have been also
derived by Capovilla and Guven, \cite{GuvenJPA02} from which we seem
to arrive at a paradox for fluid membranes: we have mentioned that
fluid membranes cannot withstand in-plane shear strain, however
Eqs.~(\ref{stress1}) and (\ref{stress2}) reveals shear stress still
exhibits in non-spherical vesicles.

\subsubsection{Solid shells}
A solid shell is a 2D isotropic continuum which can endure both
bending and in-plane shear strain. The free energy density, $G$, of
solid shells should be invariant under the in-plane coordinate
transformation. There are only two fundamental geometric invariants,
$2H$ and $K$, and two fundamental strain invariants: the trace,
$2J$, and the determinate, $Q$, of the in-plane strain tensor. Thus
free energy density should be a function of $2H$, $K$, $2J$, and
$Q$. That is, $G=G(2H,K;2J,Q)$.

If the solid shell has no initial strains and consists of materials
distributing symmetrically with regard to the middle surface of the
shell, we can expand $G$ up to the second order terms of curvatures
and strains as
\begin{equation}G=(k_c/2)(2H)^2-\bar{k}K+(k_d/2)(2J)^2-\tilde{k}Q,\label{edstyshell}\end{equation}
where $k_c$ and $\bar{k}$ are the bending moduli while $k_d$ and
$\tilde{k}$ are the in-plane rigidity moduli. The theory based on
the above free energy density is called Kirchhoff's linear shell
theory. \cite{Love44} Especially, if the shell consists of 3D
isotropic materials, we have
\begin{eqnarray}&&k_c=Yh^3/12(1-\nu^2),\label{ratiokbkc0}\\
&&k_d=Yh/(1-\nu^2),\\
&&\bar{k}/k_c=\tilde{k}/k_d=(1-\nu),\label{ratiokbkc}
\end{eqnarray} where $Y$ and $\nu$ are the
Young's modulus and Poisson ratio while $h$ is the thickness of the
shell.\cite{landau}

For a closed shell, its free energy is expressed as
Eq.~(\ref{frengFM}) with $G$ in Eq.~(\ref{edstyshell}). Of course,
we can obtain the equations of in-plane strains and shapes through
the variational method in Appendix B. The final results are the same
as those obtained from Eqs.~(\ref{forcb2D1}) and (\ref{momb2D1})
with a complement constitutive relations (\ref{consrelation2}) and
\begin{equation}\mathfrak{S}=\mathfrak{S}^i+\mathfrak{S}^f\label{consrelation3}\end{equation}
with
\begin{equation}\mathfrak{S}^i\equiv G_{\varepsilon_{11}}\mathbf{e}_{1}\mathbf{e}_{1}+(G_{\varepsilon_{12}%
}/2)( \mathbf{e}_{1}\mathbf{e}_{2}+\mathbf{e}_{2}\mathbf{e}_{1})
+G_{\varepsilon_{22}}\mathbf{e}_{2}\mathbf{e}_{2},\label{consrelation4}\end{equation}
where $G_{\varepsilon_{11}}$, $G_{\varepsilon_{12}}$, and
$G_{\varepsilon_{22}}$ represent the partial derivatives of $G$ with
respect to $\varepsilon_{11}$, $\varepsilon_{12}$, and
$\varepsilon_{22}$, the components of the in-plane strain tensor
$\mathfrak{E}$. Substituting Eq.~(\ref{edstyshell}) into
Eqs.~(\ref{consrelation2}) and (\ref{consrelation4}), and then
employing Eqs.~(\ref{forcb2D1}) and (\ref{momb2D1}), we obtain
\begin{eqnarray}
d(\mathfrak{S}_{11}^{i}\omega_{2}-\mathfrak{S}_{12}^{i}\omega
_{1})-(\mathfrak{S}_{21}^{i}\omega_{2}-\mathfrak{S}_{22}%
^{i}\omega_{1})\wedge\omega_{21}=0,&&\label{inpleq1}\\
d(\mathfrak{S}_{21}^{i}\omega
_{2}-\mathfrak{S}_{22}^{i}\omega_{1})-(
\mathfrak{S}_{11}^{i}\omega_{2}-\mathfrak{S}_{12}^{i}\omega _{1})
\wedge\omega_{12}=0,&&\label{inpleq2}
\end{eqnarray}
and {\small\begin{equation}p+2k_{c}[2H( H^{2}-K)
+\nabla^{2}H]-4(k_{d}-\tilde{k})
JH-\tilde{k}\mathfrak{R}:\mathfrak{E}=0,\label{shellshapeq}\end{equation}}where
$\mathfrak{S}_{11}^{i}=(2k_{d}J-\tilde{k}\varepsilon_{22})$,
$\mathfrak{S}_{12}^{i}=\mathfrak{S}_{21}^{i}=\tilde{k}\varepsilon
_{12}$, and
$\mathfrak{S}_{22}^{i}=(2k_{d}J-\tilde{k}\varepsilon_{11})$ are the
components of tensor $\mathfrak{S}^{i}$. $\mathfrak{R}$ is the
curvature tensor related to Eq.~(\ref{omega13}). The above equations
(\ref{inpleq1})--(\ref{shellshapeq}) describe the in-plane strains
and shapes of solid shells at equilibrium state. The similar
equations and the corresponding dynamics forms have been derived
through the variational method in Refs.~\onlinecite{TuJPA04} and
\onlinecite{Sodergaard07}, respectively, with the aid of moving
frame method.

The above equations (\ref{inpleq1}) and (\ref{inpleq2}) can be
written as one vector equation by introducing a displacement vector
$\mathbf{u}=u_1\mathbf{e}_1+u_2\mathbf{e}_2+u_3\mathbf{e}_3$, which
is related to two invariants $2J$ and $Q$ of the in-plane strain
tensor as
\begin{eqnarray}&&2J=\mathrm{div\,} \mathbf{u}-2Hu_3 \label{JQandU1}\\
&&2Q=(\mathrm{div\,} \mathbf{u}-2Hu_3)^2+(1/2)(\mathrm{curl\,}
\mathbf{u})^2-(\diamondsuit\mathbf{u})^2,\qquad\label{JQandU2}\end{eqnarray}
where
$\diamondsuit\mathbf{u}=\nabla\mathbf{u}-\mathbf{e}_3(\mathbf{e}_3\cdot\nabla\mathbf{u})$
is the in-plane part of $\nabla\mathbf{u}$. Using the new variable
$\mathbf{u}$, Eqs.~(\ref{inpleq1}) and (\ref{inpleq2}) can be be
written as
\begin{equation}(\tilde{k}-2k_{d}) \nabla (\mathrm{div\,}\mathbf{u}%
-2Hu_{3}) -\tilde{k}( \diamondsuit
^{2}\mathbf{u}+K\bar\mathbf{u}+\tilde{
\nabla}u_{3})=0,\label{2Dcauchyeq}\end{equation} where
$\bar\mathbf{u}$ and $\diamondsuit ^{2}\mathbf{u}$ are the in-plane
components of $\mathbf{u}$ and
$\mathrm{div\,}(\diamondsuit\mathbf{u})$, respectively. $\tilde{
\nabla}$ is called the gradient operator of the second class, which
is shown in our previous work.\cite{TuJPA04} In particular, $H$,
$K$, $\tilde{ \nabla}u_{3}$ vanish and $\diamondsuit ^{2}$
degenerates into $\nabla^2$ for a flat manifold. Then the above
equation degenerates into the Cauchy equation \cite{Love44} in 2D
plane. Thus Eq.~(\ref{2Dcauchyeq}) can be regarded as the Cauchy
equation in a curved surface.

\section{Application of Elastic theory in bio-structures\label{SecApBioStuct}}
In the above section, we have described fundamentals of geometric
and elastic theory on low-dimensional continua. Can this theory be
applied to the bio-structures, such as DNA and cell membranes, and
so on? DNA is a long chain macromolecule which may be described as
an elastic rod. A cell membrane is a thin structure whose thickness
and the size of the microscopic components are so much smaller than
its lateral dimension that it can be regarded as a 2D continuum
phenomenologically. We will discuss the application of the above
elastic theory in short DNA rings, lipid membranes and cell
membranes in this section.

\subsection{Short DNA ring}
DNA is a double helical structure whose diameter is about 2.5~nm.
Its bending rigidity, described as the persistence length $l_p$, is
about 50~nm (150 bp) at the room temperature. The normal DNA is
usually flexible enough because its length is so much larger than
$l_p$ that the fluctuations are quite evident. Thus the rod theory
cannot directly be applied to the normal DNA. The statistical theory
combining the rod theory is
required,\cite{MarkoSiggia95,HaijunPRL99} which is out of our topic
in this review. However, there is a special kind of short DNA rings
\cite{HanWN97,HanPNAS97,WidomMC04} which are in the length scale of
$l_p$ so that the fluctuation effect can be neglected. The diameter
is still much smaller than the total length. Thus the rod theory
mentioned in Sec.~\ref{Sectionrod} is expected to be available for
this kind of DNA rings.

Han \textit{et al.} have used AFM to observe DNA rings consisting of
several segments connected by kinks in the presence of Zn$^{2+}$
ions.\cite{HanWN97,HanPNAS97} Zhao \textit{et al.} have analyzed the
mechanism of this kink instability based on Helfrich rod theory.
\cite{Zhaow98} Their main ideas are sketched as follows. First, a
circle is a solution to Eqs.~(\ref{HfRod1}) and (\ref{HfRod2}).
Next, through analyzing the stability of the cycle, it is found
that, for the given elastic constants, there exists a critical
radius above which DNA circles will be instable. This prediction is
in good agreement with the experiments,\cite{HanWN97,HanPNAS97}
where kink deformations were observed in DNA rings of 168 bp but not
126 bp. Above some thresholds of the chiral modulus, $k_3$ in
Eq.~(\ref{FrenHlfRod}), the DNA circles turn into elliptical,
triangular, square, or other polygonal shapes, respectively. This
fact agrees with the experiments if $k_3$ is positively correlated
to the condensation of Zn$^{2+}$ ions.

Interestingly, Zhou and Ou-Yang proposed another interpretation
based on the dynamic instability of Kirchhoff rod theory
\cite{ZhouJCP99} with $\bar{\kappa}_2=\bar{\kappa}_3=0$ in
Eq.~(\ref{K-rod1}). Their result is the same as that obtained
directly from the first and second order variations of the free
energy. We deal with the latter scenario. First, $\tau=0$, $\phi=0$,
and $\kappa=1/R$ satisfy Eqs.~(\ref{RodShape1})--(\ref{RodShape3})
derived from the first order variation of the free energy. That is,
a planar circle with radius $R$ is an equilibrium configuration.
Next, through the second order variation of the free energy, we can
obtain the characteristic function describing the stability of the
circle
\begin{equation}g_c(R)=\bar{\kappa}_1^2-(1-\Gamma)\bar{\kappa}_1/R-\Gamma n^2/R^2\leq 0\end{equation}
where $n>1$ is an arbitrary integer and $\Gamma=k_3/k_1$. From the
above inequality, we obtain the critical radius
\begin{equation}R_c=8\Gamma/\bar{\kappa}_1[\Gamma-1+\sqrt{(\Gamma-1)^2+16\Gamma}],\end{equation}
above which the circle is instable. If only the presence of
Zn$^{2+}$ ions tunes the values of $\Gamma$ and $k_3/k_2$ such that
$R_c$ is in the range between $63/\pi$ (bp) and $84/\pi$ (bp), the
above result is also in agreement with the
experiments,\cite{HanWN97,HanPNAS97} where kink deformations were
observed in DNA rings of 168 bp but not 126 bp.

\begin{figure}[pth!]
\includegraphics[width=5cm]{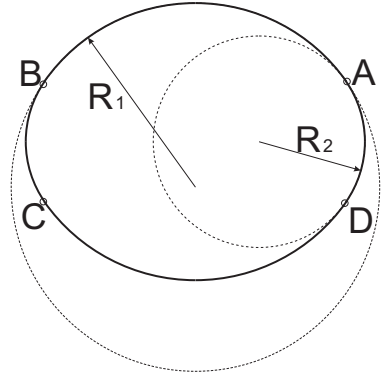}\caption{\label{figdouble2arc}
A possible configuration of a short DNA ring.}
\end{figure}

In Sec.~\ref{Sectionrod}, we also mention the theory of
bending-soften rod. Can this theory also provide an interpretation
to the experiments? Let us consider the bending-soften rod theory of
the first kind whose free energy density is expressed as
Eq.~(\ref{BSRodFEG1}). When the radius $R$ of the ring is smaller
than $1/\kappa_c$, any small perturbation will increase the free
energy. If $R>1/\kappa_c$, the ring might transform into the
fictitious configuration shown in Fig.~\ref{figdouble2arc} which
consists of four arcs AB, BC, CD, DA with the radius $R_1$ and
$R_2$. To see conveniently, the joint points are marked as small
cycles in the figure. Obviously, $R_2<R<R_1$. Through simple
calculations, we find that the fictitious configuration is
energetically less favorable than the perfect ring with radius $R$.
Therefore, this coarse analysis reveals that the theory of
bending-soften rod cannot explain the experiments.

\subsection{Lipid membrane\label{Seclipm}}
\begin{figure}[pth!]
\includegraphics[width=5.5cm]{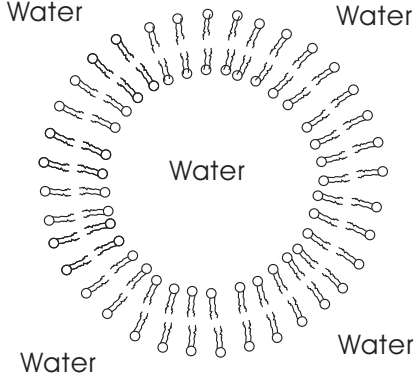}\caption{\label{figvesicleb}
A lipid bilayer vesicle.}
\end{figure}

Lipids are dominant composition of cell membranes. Most of lipid
molecules have a polar hydrophilic head group and two hydrophobic
hydrocarbon tails. When a quantity of lipid molecules disperse in
water, they will assemble themselves into a bilayer vesicle as
depicted in Fig.~\ref{figvesicleb}, in which the hydrophilic heads
shield the hydrophobic tails from the water surroundings because of
the hydrophobic forces. This self-assembly process has been
numerically investigated by Lipowsky \emph{et
al.}\cite{GoetzJCP98,GoetzPRL99,ShillcockJPC06} and Noguchi \emph{et
al.}\cite{NoguchiPRE06} through molecular dynamics simulation based
on coarse-grained model or meshless membrane model.

The thickness of the lipid bilayer and the size of single lipid
molecules are much smaller than the scale of the whole lipid
bilayer. Additionally, at the physiological temperature, the lipid
bilayer is usually at the nematic state where the hydrocarbon chains
of the lipid molecules are roughly perpendicular to the bilayer
surface. Thus the bilayer can be regarded as a 2D fluid membrane
whose free energy density is expressed as Eq.~(\ref{freeng-FM}).
Expanding it up to the second order terms of curvatures, we obtain
the Helfrich's form:\cite{Helfrich73}
\begin{equation}
G_H=(k_c/2)(2H+c_0)^2-\bar{k}K+\lambda,\label{HelfrichBL}
\end{equation}
where $k_c$ and $\bar{k}$ are the bending moduli of the lipid
bilayer. We emphasize that the minus sign before $\bar{k}$ in
Eq.~(\ref{HelfrichBL}) is opposite to Helfrich's convention.
$\lambda$ is the surface tension of the bilayer. $c_0$ is called the
spontaneous curvature that reflects asymmetric factors between two
sides of the bilayer, including the lipid distribution, the chemical
environment, and so on. $k_c$ is about 20 $\mathrm{T}$ for lipid
bilayers, where the Boltzmann factor is set to 1 and $\mathrm{T}$
the room temperature, from which the persistence length of lipid
bilayers is estimated about 10 $\mu$m.\cite{Lipowsky91,Seifert97} In
this section we only consider the size of lipid bilayers smaller
than 10 $\mu$m so that the fluctuation effect on the shape of lipid
bilayers can be neglected. The model based on Eq.~(\ref{HelfrichBL})
is called spontaneous curvature model. We still remind gentle
readers to note the two similar nonlocal models--- the
bilayer-coupling model \cite{Svetina83,Seifert91} and the area
difference model,\cite{LingMiao94} although we will not touch them
in the present review.

\subsubsection{Closed vesicles}
The free energy of a lipid vesicle under the osmotic pressure $p$
(the outer pressure minus the inner one) can be written as
Eq.~(\ref{frengFM}) with $G=G_H$ being Helfrich's form
(\ref{HelfrichBL}). Substituting (\ref{HelfrichBL}) into
Eq.~(\ref{ELclosFM}), we can obtain the shape equation of lipid
vesicles: \cite{OYPRL87,OYPRA87} {\small\begin{equation}p-2\lambda
H+k_c(2H+c_0)(2H^2-c_0H-2K)+2k_c\nabla^2H=0.\label{shape-closed}\end{equation}}This
equation is the fourth order nonlinear equation. It is not easy to
find its special solutions. We have known three typical analytical
solutions: sphere, \cite{OYPRL87} torus,\cite{oypra90,Seiferttorus}
and biconcave discoid shape. \cite{NaitoPRE93}

For a sphere with radius $R$, we have $H=-1/R$ and $K=1/R^2$.
Substituting them into (\ref{shape-closed}), we arrive at
\begin{equation}pR^2+2\lambda
R-k_cc_0(2-c_0R)=0.\label{sphericalbilayer}\end{equation} This
equation gives the sphere radius under the osmotic pressure $p$.

A torus is a revolution surface generated by a circle with radius
$\rho$ rotating around an axis in the same plane of the circle. The
revolution radius $r$ should be larger than $\rho$. A point in the
torus can be expressed as a vector
$\{(r+\rho\cos\varphi)\cos\theta,(r+\rho\cos\varphi)\sin\theta,\rho\sin\varphi\}$.
Through simple calculations, we have $2H=-(r+2\rho\cos\varphi)/\rho(
r+\rho\cos\varphi)$, $K=\cos\varphi/\rho(r+\rho\cos\varphi)$.
Substituting them into Eq.~(\ref{shape-closed}), we derive
\begin{eqnarray}
&&\hspace{0.26cm} [(2 k_c c_0^2\rho^2-4 k_c c_0\rho+4\lambda
\rho^2+2P\rho^3)/{\varrho^3}]\cos^3\varphi\nonumber\\
&&+[(5 k_c c_0^2\rho^2-8 k_c c_0\rho+10\lambda \rho^2+6P\rho^3)/{\varrho^2}]\cos^2\varphi\nonumber\\
&&+[{(4 k_c c_0^2\rho^2-4 k_c c_0\rho+8\lambda \rho^2+6P\rho^3)}/{\varrho}]\cos\varphi\nonumber\\
&&+2k_c/{\varrho^2}+k_c(c_0^2\rho^2-1)+2(P\rho+\lambda)\rho^2=0,\label{toruseq}
\end{eqnarray}
where $\varrho=r/\rho$. If $\varrho$ is finite, then
Eq.~(\ref{toruseq}) holds if and only if the coefficients of
$\{1,\cos\varphi,\cos^2\varphi,\cos^3\varphi\}$ vanish. It follows
$2\lambda \rho^{2}=k_{c}c_{0}\rho( 4-c_{0}\rho)$,
$P\rho^{3}=-2k_{c}\rho c_{0}$ and $\varrho=\sqrt{2}$.\cite{oypra90}
That is, there exists a lipid torus with the ratio of its two
generated radii being $\sqrt{2}$, which was confirmed in the
experiment \onlinecite{MutzPRA91}.

To describe the solution of biconcave discoid shape, we write the
shape equation (\ref{shape-closed}) under the axisymmetric
condition. If a planar curve $z=z(\rho)$ revolves around the
$z$-axis, an axisymmetric surface is formed. Each point on the
surface is expressed as $\mathbf{r}=\{\rho\cos \varphi, \rho\sin
\varphi, z(\rho)\}$. Denote $\psi=\arctan(dz/{d\rho})$ and
$\Psi=\sin\psi$. Then Eq.~(\ref{shape-closed}) is transformed into
\cite{HuJGPRE93} {\small\begin{eqnarray}
\frac{1}{2}\left[ \frac{\left( \rho \Psi \right) ^{\prime }}{%
\rho }+c_{0}\right] \left\{ \left[ \rho \left( \frac{\Psi }{\rho
}\right)
^{\prime }\right] ^{2}-\frac{c_{0}\left( \rho \Psi \right) ^{\prime }}{\rho }%
\right\}-\frac{\lambda \left( \rho \Psi \right) ^{\prime }}{%
k_{c}\rho }  &&\nonumber\\
+\left\{ \rho \left[ \frac{\left( \rho \Psi \right) ^{\prime }}{\rho
}\right]
^{\prime }\right\} ^{\prime }\frac{1-\Psi ^{2}}{\rho }-\left[ \frac{%
\left( \rho \Psi \right) ^{\prime }}{\rho }\right] ^{\prime }\Psi
\Psi ^{\prime
}+\frac{p}{k_{c}}=0,&&\label{nequilbcl}\end{eqnarray}}where the
prime represents the derivative with respect to $\rho$. This
equation is called the shape equation of axisymmetric lipid
vesicles. Its first integral, group structure and corresponding
Hamilton's equations are investigated by Zheng and Liu,
\cite{ZhengWMPRE93} Xu and Ou-Yang,\cite{XuOY04} and Capovilla et
al.\cite{GuvenJPA051,GuvenJPA052} respectively.

\begin{figure}[pth!]
\includegraphics[width=5.5cm]{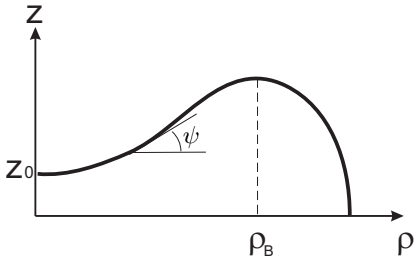}\caption{\label{figbiconcave}
A quarter outline of the biconcave surface.}
\end{figure}

It is easy to verify that $\Psi=\sin\psi=-c_0\rho\ln(\rho/\rho_B)$
with a constant $\rho_B$ is a solution to Eq.~(\ref{nequilbcl}) if
$p$ and $\lambda$ are vanishing. For $0<c_0\rho_B<e$, the parameter
equation
\begin{equation}\left\{\begin{array}{l}\sin\psi=-c_0\rho\ln(\rho/\rho_B)\\
z=z_0+\int_0^\rho \tan\psi d\rho
\end{array}\right.\label{solutionbicon}\end{equation}
corresponds to a curve shown in Fig.~\ref{figbiconcave}. A biconcave
discoid surface will be achieved when this curve revolves around
$z$-axis and then reflects concerning the horizontal plane. The
above equation (\ref{solutionbicon}) can give a good explanation to
the shape of human red blood cell under normal physiological
conditions. \cite{NaitoPRE93} If $c_0\rho_B$ is out of the range
between 0 and $e$, Eq.~(\ref{solutionbicon}) corresponds to a
prolate ellipsoid or other self-intersecting
surfaces.\cite{LiuqhPRE99}

In the purely mathematical viewpoint, there are also the other
solutions to Eq.~(\ref{shape-closed}) such as cylinder, constant
mean curvature surface, periodic undulation
surface,\cite{ZhangSGPRE96} pearling tubule, \cite{Mladenov02} and
so on.\cite{oybook,GLandolfi} However, It is a pity that they are
open surfaces and do not correspond to truly closed vesicles.

As mentioned above, it is fairly difficult to find the analytical
solution to Eq.~(\ref{shape-closed}). Thus we appreciate the
applications of numerical methods to find the equilibrium shapes of
closed vesicles. Two kinds of typical numerical frameworks are
usually employed. The first one is to use `Surface Evolver', a
software package developed by Brakke,\cite{Brakkeexpm} to find the
configurations minimizing the free energy under some
constraints.\cite{YanjPRE98,ZhouJMPB01,ZhangJMPB02,ZhangSG06} The
second one is based on the phase field formulation of Helfrich's
free energy density (\ref{HelfrichBL}) and diffusive interface
approximation.\cite{DuJCP04,DuJCP05,DuCPAA05,DuLiuWang06} The above
numerical methods can obtain lipid vesicles with different shapes
either axisymmetric or asymmetric. Additionally, the finite element
method might be a potential method although very sparse literature
\cite{FengKlug06} treats lipid bilayers by using it.

\subsubsection{Stability of closed vesicles}
When the osmotic pressure is beyond some threshold, a closed vesicle
will lose its stability and change its shape abruptly. The threshold
is called the critical pressure. To obtain it, one should calculate
the second order variation of the free energy (\ref{frengFM}) with
$G$ being Helfrich's form (\ref{HelfrichBL}), which has been dealt
with in the general case as: \cite{TuJPA04,CapovillaJPA04}
{\small\begin{eqnarray}&&\hspace{-0.3cm}\delta^2 \mathcal{F}=\int
k_{c}[ (\nabla ^{2}\Omega _{3})^{2}+(2H+c_{0})\nabla (2H\Omega
_{3})\cdot \nabla \Omega _{3}] dA\nonumber
\\
&&\hspace{-0.3cm}+\int [4k_{c}(2H^{2}-K)^{2}+k_{c}K(
c_{0}^{2}-4H^{2}) +2\lambda
K-2Hp]\Omega _{3}^{2}dA\nonumber\\
&&\hspace{-0.3cm}+\int
[k_{c}(14H^{2}+2c_{0}H-4K-c_{0}^{2}/2)-\lambda ]\Omega _{3}\nabla
^{2}\Omega _{3}dA \nonumber\\
&&\hspace{-0.3cm}-2k_{c}\int (2H+c_{0})[\nabla \Omega _{3}\cdot
\tilde{\nabla}\Omega _{3}+2\Omega _{3}\nabla \cdot
\tilde{\nabla}\Omega _{3}]dA,\label{secondvlipidb}
\end{eqnarray}}where $\Omega _{3}$ is an arbitrary
small out-of-plane displacement and the operator $\tilde{\nabla}$ is
the gradient operator of the second class.\cite{TuJPA04}

Here we will mention two results for special configurations.

First, let us consider a lipid sphere that satisfies
Eq.~(\ref{sphericalbilayer}). On the sphere, the function $\Omega
_{3}$ can be expanded by the spherical harmonic functions $Y_{lm}$
as $\Omega _{3} =\sum_{l=0}^{\infty }\sum_{m=-l}^{m=l}a_{lm}Y_{lm}$.
Substituting it into Eq.~(\ref{secondvlipidb}), we derive
$\delta^{2}\mathcal{F}=
(R/2)\sum_{l,m}|a_{lm}|^{2}[l(l+1)-2]\{2k_{c}[l(l+1)-c_{0}R]/R^{3}-p\}$,
From which we can obtain the critical pressure\cite{OYPRL87}
\begin{equation}p_c=2k_{c}(6-c_{0}R)/R^{3}.\label{criticalpsc}\end{equation}
If $p<p_c$, $\delta^{2}\mathcal{F}\geq 0$ for any $|a_{lm}|$; on the
contrary, $\delta^{2}\mathcal{F}$ can be negative for the special
selection of $|a_{lm}|$. The above equation depends also on $c_0$.
If $c_0>6/R$, then $p_c$ is negative, which reveals that a sphere
vesicle is always instable for large enough $c_0$.

Next, let us still regard a long enough lipid tubule as a closed
vesicle. Denoted its radius as $\rho$. From Eq.~(\ref{shape-closed})
we have \begin{equation}( k_{c}/2) (1/\rho
^{2}-c_{0}^{2})-p\rho=\lambda.\label{radiusliptube}\end{equation} On
the cylindrical surface, $\Omega _{3}$ can be expanded as Fourier
series $\Omega _{3}=\sum_{l=-\infty}^{\infty }a_l\exp(il\theta)$.
Substituting it into Eq.~(\ref{secondvlipidb}) and combining
Eq.~(\ref{radiusliptube}), we derive
$\delta^{2}\mathcal{F}=\sum_{l=-\infty}^{\infty } |a_{l}|^{2}(
l^{2}-1) [k_{c}( l^{2}-1) /\rho ^{3}-p]$, From which we can obtain
the critical pressure
\begin{equation}p_c=3k_{c}/\rho^{3}.\label{criticalptube}\end{equation}
If $p<p_c$, $\delta^{2}\mathcal{F}\geq 0$ for any $|a_{l}|$; on the
contrary, $\delta^{2}\mathcal{F}$ can be negative for the special
selection of $|a_{l}|$.

\subsubsection{Open vesicles with free edges}
The opening-up process of lipid vesicles by talin, a protein, has
recently been observed \cite{Saitoh,Nomura} which pushes us to study
the equilibrium equation and boundary conditions of lipid vesicles
with free exposed edges. Capovilla \textit{et al.} have addressed
this problem and given the equilibrium equation and boundary
conditions.\cite{GuvenPRE02} Inspired by the talk ``moving frame
method'' of Chern,\cite{Chern} we introduce exterior differential
form to deal with the variational problem on open surface and obtain
concisely the shape equation and boundary conditions of open lipid
vesicles.\cite{TuPRE03} Numerical solution to the shape equation and
boundary conditions with relaxed method can explain the experimental
results very well. \cite{HotaniPRE05} A quantity of open vesicles
with free edges have also been obtained numerically by Wang and Du
\cite{WangDu06} with the phase field method. Here we will not
further discussed the dynamical opening process of the vesicles,
which has been recently investigated by Kaga and
Ohta.\cite{KagaOhta}

\begin{figure}[pth!]
\includegraphics[width=5.5cm]{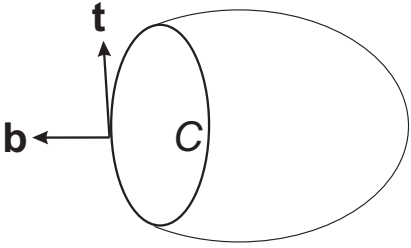}\caption{\label{figopensurface}
An open surface with boundary curve $C$.}
\end{figure}

We regard an open lipid vesicle with a free edge as a smooth surface
with a boundary curve $C$, as shown in Fig.~\ref{figopensurface}.
$\mathbf{t}$ is the tangent vector of the curve $C$. $\mathbf{b}$,
in the tangent plane of the surface, is perpendicular to
$\mathbf{t}$ and points to the opposite side that the surface
located in. The free energy of the open lipid vesicle is written as
\begin{equation}
\mathcal{F}=\int G_H\, dA+\gamma \oint_{C}ds,\label{FrengopenV}
\end{equation}
where $\gamma$ represents the line tension of the edge and $G_H$ has
the Helfrich's form (\ref{HelfrichBL}).

The first order variation of $\mathcal{F}$ gives the shape equation
\begin{equation}k_{c}(2H+c_{0})(2H^{2}-c_{0}H-2K)-2\lambda H+2k_{c}\nabla ^{2}H =0,
\label{shapeopenLV}\end{equation} and the boundary conditions
as:\cite{TuPRE03}
\begin{eqnarray}
&&\left[ k_{c}(2H+c_{0})-\bar{k}k_n\right]_{C} =0,\label{openbound1} \\
&&\left[ 2k_{c}{\partial H}/{\partial\mathbf{b}}+\gamma
k_n-\bar{k}\tau_g'\right]_{C} =0,\label{openbound2}\\
&&\left[ G_H +\gamma k_{g}\right]_{C}=0,\label{openbound3}
\end{eqnarray}
where $k_n$ and $k_g$ are normal curvature and geodesic curvature of
the boundary curve $C$. $\tau_g'$ is the derivative of geodesic
torsion $\tau_g$ with respect to the arc length of curve $C$. The
mechanical meanings of the above four equations are as follows:
Eq.~(\ref{shapeopenLV}) is the normal force balance equation of the
membrane; Eq.~(\ref{openbound1}) is the moment balance equation of
points in curve $C$ around the direction of $\mathbf{t}$;
Eq.~(\ref{openbound2}) is the force balance equation of points in
curve $C$ along the normal direction of surface; and
Eq.~(\ref{openbound3}) is the force balance equation of points in
curve $C$ along the direction of $\mathbf{b}$. It is necessary to
emphasize that the boundary conditions are available for open
vesicles with more than one free edge because the edge in our
derivation is a general one.

In Ref.~\onlinecite{TuPRE03}, we have shown two analytical solutions
to above equations (\ref{shapeopenLV})--(\ref{openbound3}): One is a
cup-like membrane and another is the central part of a torus.
Several numerical solutions to these equations are obtained by Umeda
\textit{et al.}\cite{HotaniPRE05}. Their results reveal that the
line tension $\gamma$ induced by talin correlates negatively with
the concentration of talin, which is in agreement with the
experimental result that the hole of vesicle is enlarged with the
concentration of talin.\cite{Saitoh}

\subsubsection{Vesicles with lipid domains}
The above discussion on open lipid vesicles with free edges can be
extended to study a vesicle of several lipid components. The domains
usually formed so that each domain contains one or two kinds of
lipid molecules. The morphology of axisymmetric vesicles with
multi-domains has been theoretically investigated by J\"{u}licher
and Lipowsky. \cite{Lipowsky93} It is found that lipid domains
facilitate the budding of vesicles.\cite{LipowskyJPC03} The giant
vesicles with lipid domains have been observed in recent
experiment.\cite{Baumgart03} There are two kinds of lipid domains
which are at the liquid-ordered state and liquid-disordered state,
respectively. It is natural to assume that different kinds of
domains have different bending moduli and spontaneous curvatures.
The axisymmetric vesicles in the experiment can be explained with
J\"{u}licher-Lipowsky theory through numerically method. Baumgart
\emph{et al.} have demonstrated that the line tension, the osmotic
pressure, the relative bending moduli, and the spontaneous curvature
have significant effects on the morphology of a vesicle with two
domains being at the liquid-ordered and disordered states,
respectively.\cite{Baumgart05}

\begin{figure}[pth!]
\includegraphics[width=5.5cm]{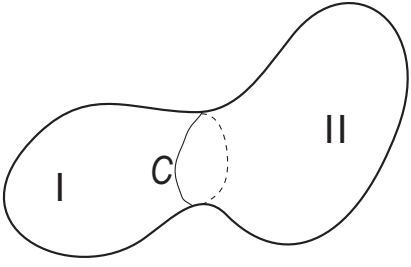}\caption{\label{figlipdomain}
A vesicle with two domains separated by curve $C$.}
\end{figure}

The asymmetric vesicles are also experimentally observed in
Ref.~\onlinecite{Baumgart03}, which enlightens us to investigate the
shape equation of each domains and the boundary conditions between
domains without any axisymmetric assumptions. Let us consider a
vesicle with two domains separated by curve $C$ sketched in
Fig.~\ref{figlipdomain}. The free energy can be expressed
as\cite{Lipowsky93}
\begin{equation}\mathcal{F}=\int G_H^I\, dA + \int G_H^{II}\,dA + \gamma\oint ds+ p\int dV,\label{freeVDM}\end{equation}
where $G_H^I$ and $G_H^{II}$ have the Helfrich from
(\ref{HelfrichBL}) with the bending moduli $k_c^I$, $\bar{k}^I$,
$k_c^{II}$, $\bar{k}^{II}$, the spontaneous curvatures $c_0^I$,
$c_0^{II}$, and the surface tensions $\lambda^I$, $\lambda^{II}$,
respectively. The integrals in the first and second terms of
Eq.~(\ref{freeVDM}) are performed on the domain I and II shown in
Fig.~\ref{figlipdomain}, respectively. $\gamma$ is the line tension
of boundary curve $C$. $p$ is the osmotic pressure of the vesicle.

In terms of the physical meanings of
Eqs.~(\ref{shapeopenLV})--(\ref{openbound3}), we can easily write
down the shape equation of domains as:\cite{TuJPA04,TuZCTSF}
{\small\begin{equation}p-2\lambda^i
H+k_c^i(2H+c_0)(2H^2-c_0^iH-2K)+2k_c^i\nabla^2H=0.\label{shape-domain}\end{equation}}where
the superscript $i=$ I and II represents the physical quantity of
lipid domains I and II, respectively. Additionally, the boundary
conditions between domains are as follows:\cite{TuJPA04,TuZCTSF}
\begin{eqnarray}&&\hspace{-0.9cm}\lbrack k_{c}^{I}(2H+c_{0}^{I})-k_{c}^{II}(2H+c_{0}^{II})-(\bar{k}^{I}-\bar{k%
}^{II})k_{n}\rbrack_{C} =0,\label{boundtwocom1}\\
&&\hspace{-0.9cm}\lbrack 2(k_{c}^{I}+k_{c}^{II})\partial H/\partial \mathbf{b}-(\bar{k}^{I}+%
\bar{k}^{II})\tau_g'+\gamma k_{n}\rbrack_{C} =0,\label{boundtwocom2}\\
&&\hspace{-0.9cm}\lbrack G^I-G^{II} +\gamma k_{g}\rbrack_C
=0,\label{boundtwocom3}\end{eqnarray} where $\mathbf{b}$ is
perpendicular to the boundary curve $C$ and points to the side of
domain II.

As we know, there is still no any numerical result on asymmetric
vesicles with domains directly from the above equations in the
previous literature. Only in Ref.~\onlinecite{WangDu06}, Wang and Du
discussed the morphology of asymmetric vesicles with domains through
the phase field model.

In the above theory, the detailed architecture of liquid-ordered and
disordered phases is neglected. There are special lipid domains at
liquid-ordered phase, so called rafts, which are enriched in
cholesterol and sphingolipids.\cite{SimonsNat97} Cholesterol is a
kind of chiral molecules, which has not been included in the above
theory. Recently, a concise theory of chiral lipid membranes
developed by Tu and Seifert\cite{TuSeifert1} might be extended to
discuss the raft domains.

\subsubsection{Adhesions of Vesicles}
Cell adhesion is a complex biological process which controls many
functions of life. It can be understood as a first-order wetting
transition\cite{Sackmann02CPC} and might be simplified as the
adhesion of lipid vesicles. As a model, Seifert and Lipowsky have
theoretically investigated a lipid vesicle adhering to a flat rigid
substrate and found that the vesicle undergoes a nontrivial adhesion
transition from the free state to the bound state, which is governed
by the competition between the bending and adhesion
energies.\cite{Seifert90} Ni \textit{et al.} have discussed the
adhering lipid vesicles with free edges and the adhesion between a
lipid tubule with a rigid substrate.\cite{NiCSB05,NiIJMPB06} A big
progress on this topic is recently made by Guven and his
coworkers\cite{CapovillaPRE02,DesernoCM07} who obtain the general
equations to describe the contact line between the vesicle and the
rigid substrate or another vesicle.

\begin{figure}[pth!]
\includegraphics[width=7.5cm]{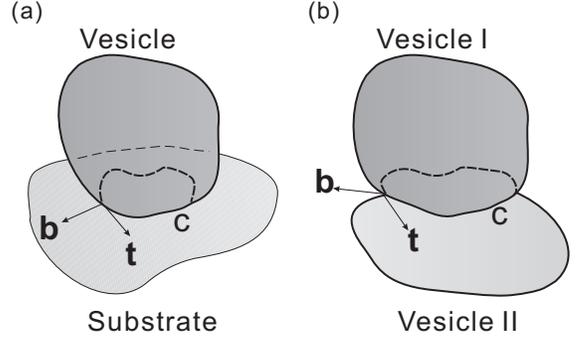}\caption{\label{figadhesionves}
Adhesions. (a) Adhesion between a lipid vesicle and rigid substrate
with a contact line C. (b) Adhesion between two lipid vesicles with
a contact line C.}
\end{figure}

The adhesion between a lipid vesicle and a rigid substrate is
depicted in Fig.~\ref{figadhesionves}a where the contact area is
denoted by $\bar{A}$. The free energy of this system is expressed
as\cite{Seifert90}
\begin{equation}\mathcal{F}=\int G_H\, dA + p\int dV-W\bar{A},\label{fregadhesion}\end{equation}
where $p$ is the osmotic pressure of the vesicle and $W$ is the
strength of the adhesion potential between the vesicle and the
substrate. $G_H$ is the free energy density of Helfrich's form
(\ref{HelfrichBL}). For the flat rigid substrate, a characteristic
radius and the length scale of the vesicle are defined as
$R_a=\sqrt{2k_c/W}$ and $R=\sqrt{A/4\pi}$, respectively. If $R<R_a$,
the vesicle is a little stiffer or the attraction is relative weak
such that $\bar{A}$ approaches to zero. Thus the vesicle is unbound
to the substrate and this state is called the free state. On the
contrary, the vesicle is at the bound state. At this state, let us
take $\mathbf{t}$ as the tangent vector of the contact line $C$, and
$\mathbf{b}$ perpendicular to $\mathbf{t}$ and in the common tangent
plane of the lipid vesicle and the substrate. The absolute value of
the normal curvature along $\mathbf{b}$ for the point on the contact
line is proven to be $\sqrt{2W/k_c}$ for an axisymmetric vesicle
adhering to the flat substrate.\cite{Seifert90} If the rigid
substrate is curved, the above conclusion is revised
as\cite{DesernoCM07}
\begin{equation}|\kappa^V_{\mathbf{b}}-\kappa^S_{\mathbf{b}}|=\sqrt{2W/k_c}\ ,\end{equation}
where $\kappa^V_{\mathbf{b}}$ and $\kappa^S_{\mathbf{b}}$ are the
normal curvatures along $\mathbf{b}$ for the points outside but near
the contact line, calculated by using the surfaces of the vesicle
and the substrate, respectively.

The adhesion between two lipid vesicles is depicted in
Fig.~\ref{figadhesionves}b. The free energy of this system is
expressed as\cite{Seifert90} {\small\begin{equation}\mathcal{F}=\int
G_H^I\, dA +\int p^I dV+\int G_H^{II}\, dA + \int p^{II}
dV-W\bar{A},\label{fregadhesion2}\end{equation}}where $p^I$ and
$p^{II}$ are the osmotic pressures of the vesicles I and II,
respectively. $\bar{A}$ and $W$ are the contact area and adhesion
strength, respectively. $G_H^I$ and $G_H^{II}$ are the Helfrich's
free energy density of vesicle I and II. The first order variation
of (\ref{fregadhesion2}) gives the same shape equation of two
vesicles as (\ref{shape-domain}) and the adhesion boundary
conditions:\cite{DesernoCM07}
\begin{eqnarray}
&&(1+k_c^I/k_c^{II})(\kappa^I_{\mathbf{b}}-\kappa^A_{\mathbf{b}})^2=2W/k_c^I,\label{BCAd1}\\
&&(1+k_c^{II}/k_c^I)(\kappa^{II}_{\mathbf{b}}-\kappa^A_{\mathbf{b}})^2=2W/k_c^{II},\label{BCAd2}\\
&&\partial(\kappa^I_{\mathbf{b}}+\kappa^{II}_{\mathbf{b}}-\kappa^A_{\mathbf{b}})/\partial\mathbf{b}=0,\label{BCAd3}
\end{eqnarray} where $\kappa^I_{\mathbf{b}}$ and
$\kappa^{II}_{\mathbf{b}}$ are the normal curvatures along
$\mathbf{b}$ for the points outside the adhesion domain but near the
contact line calculated by using the surfaces of vesicles I and II,
respectively. $\kappa^A_{\mathbf{b}}$ is the normal curvature for
the points inside the adhesion domain but near the contact line
calculated by using the common surface of vesicles I and II. As we
know, there is still lack of numerical solutions to the above
equations (\ref{BCAd1})--(\ref{BCAd3}) in the previous literature.
Only in the recent work, Ziherl and Svetina\cite{ZiherlPNAS07} have
investigated the adhesion between two vesicles by numerically
minimizing the free energy (\ref{fregadhesion2}) with
$k_c^{II}=k_c^{I}$ and various $W$.

Is the behavior of vesicle adhesion close to that of cell adhesion?
The cell membrane can bear shear strain whose adhesion behavior
might be much closer to the adhesion between a polyelectrolyte
microcapsule and the substrate.\cite{Graf06} Interestingly, beyond
the threshold adhesion strength $W_c$, the contact length scale
increases in proportion to $(W-W_c)^{1/2}$, which is the same as the
behavior of vesicle adhesions except the coefficient before
$(W-W_c)^{1/2}$.

\subsubsection{A different viewpoint of surface tension}
Although the lipid bilayer cannot withstand the in-plane shear
strain, it can still endure the in-plane compression strain. The
in-plane compression modulus, $k_b$, of lipid bilayers is about 0.24
N/m.\cite{RawiczBJ2000} Considering this point, we may write the
free energy of a closed lipid vesicle as
\begin{equation}\mathcal{F}=p\int dV+\int G_B dA+\int(k_{b}/2)(2J_b)^2dA,\label{FrengCV}\end{equation}
where
\begin{equation}G_B=(k_c/2)(2H+c_0)^2-\bar{k}K,\label{freelipid}\end{equation} and
$J_b$ is the in-plane compression or stretch strain. We emphasize
that the contribution of chemical potential are omitted when we
write the above free energy.

The first order variation of the free energy (\ref{FrengCV}) reveals
that $2J_b$ is a constant and then
\begin{eqnarray}&&p-2(2k_{b}J_b)H+2k_c\nabla^2H\nonumber\\
&&\hspace{0.3cm}+k_c(2H+c_0)(2H^2-c_0H-2K)=0.\end{eqnarray}
Comparing the above equation with the shape equation
(\ref{shape-closed}) of lipid vesicles, we deduce that
\begin{equation}\lambda=2k_bJ_b.\end{equation}

In the discussion on the stability of closed lipid vesicles, we have
seen that the surface tensor $\lambda$ has no effect on the critical
pressure. The second order variation of the free energy
(\ref{FrengCV}) can give the same conclusion. $\delta^2[p\int
dV+\int G_B dA]$ has been shown in Eq.~(\ref{secondvlipidb}) with
vanishing $\lambda$. The additional term is
\begin{equation}\delta^2 \int(k_{b}/2)(2J_b)^2dA=\int k_{b}(\mathrm{div\,} \textbf{v}-2H\Omega_3)^2dA\end{equation}
where
$\mathbf{v}=\Omega_1\mathbf{e}_1+\Omega_2\mathbf{e}_2+\Omega_3\mathbf{e}_3$
represents the infinitesimal displacement vector of the vesicle
surface. We can always select the proper deformation modes such that
$\mathrm{div\,} \textbf{v}-2H\Omega_3=0$ and then $\delta^2
\int(k_{b}/2)(2J_b)^2dA$ vanishs, but $\delta^2[p\int dV+\int G_B
dA]$ is not affected. That is, the critical pressure is determined
merely by $\delta^2[p\int dV+\int G_B dA]$, which is independent on
the compression modulus of lipid bilayer $k_b$.

\subsection{Cell membrane\label{Seccellmb}}
Cell membrane consists of lipids, proteins, and a small quantity of
carbohydrates and so on. A simple but widely accepted model for cell
membranes is the fluid mosaic model\cite{nicolson72} proposed by
Singer and Nicolson in 1972. In this model, the cell membrane is
considered as a lipid bilayer where the lipid molecules can move
freely in the membrane surface like fluid, while the proteins are
embedded in the lipid bilayer. Some proteins, so called integral
membrane proteins, traverse entirely in the lipid bilayer and play
the role of information and matter communications between the
interior of the cell and its outer environment. The others, so
called peripheral membrane proteins, are partially embedded in the
bilayer and accomplish the other biological functions. Beneath the
lipid membrane, the membrane skeleton, a network of proteins, links
with the proteins embedded in the lipid membrane. Mature mammalian
and human erythrocytes (i.e., red blood cells) are lack of a cell
nucleus. Thus they provide a good experimental model for studying
the mechanical properties of cell
membranes.\cite{EvansBJ76,EvansBJ83,EngelhardtBJ88,Lenormand} On the
theoretical side, spontaneous curvature model,\cite{Helfrich73}
rubber membrane model,\cite{Evans73,FungBJ68,EvansBJ73} and dual
network model\cite{BoalPRL92} have been employed to investigate the
mechanical and thermal fluctuation properties of erythrocyte
membranes. We will address the elasticity and stability of composite
shell model for cell membranes in this section.

\subsubsection{Composite shell model of cell membranes}
\begin{figure}[pth!]
\includegraphics[width=7.5cm]{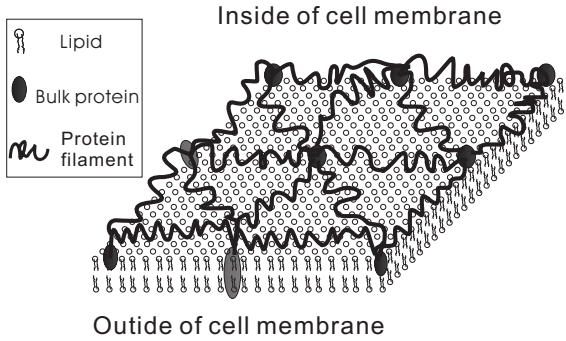}\caption{\label{figcompositmb}
Local schematic picture of the composite shell model for a cell
membrane.}
\end{figure}

A cell membrane can be simplified as a composite shell
\cite{Sackmannbook} of lipid bilayer and membrane skeleton. The
membrane skeleton, inside of the cell membrane, is a network of
protein filaments as shown in Fig.~\ref{figcompositmb}. The joint
points of the network are bulk proteins embedded in the lipid
bilayer. The whole membrane skeleton seems to float the sea of the
lipid bilayer. It can have a global movement along the surface of
the bilayer but the movement of the joints along the normal
direction is totally coupling with the bilayer. In the mechanical
point of view, the lipid bilayer can endure the bending deformation
but hardly bear the in-plane shear strain. On the contrary, the
membrane skeleton can endure the in-plane shear strain but hardly
bear the bending deformation. The composite shell overcomes the
shortage of the lipid bilayer and the membrane skeleton. It can
sustain both bending deformation and in-plane shear strain.

The contour length of protein chain between joints in the membrane
skeleton is about 100~nm which is much smaller than the size ($\sim
10\,\mu$m) of cell membranes. The lipid bilayer is 2D homogenous.
The membrane skeleton is roughly a 2D locally hexagonal lattice. As
is well known, the mechanical property of a 2D hexagonal lattice is
2D isotropic.\cite{nyebook} Thus the composite shell of the lipid
bilayer plus the membrane skeleton can still be regarded as a 2D
isotropic continuum. Its free energy density should be invariant
under the in-plane coordinate transformation and can be written as
$G_{cm}=G_{cm}(2H,K;2J,Q)$. We can expand $G_{cm}$ up to the second
order terms of curvatures and strains as
\begin{equation}G_{cm}=G_B+(k_b/2)(2J_b)^2+G_{sk},\label{edstycellm}\end{equation}
where $G_B$ results mainly from the bending energy of the lipid
bilayer, which has the form as Eq.~(\ref{freelipid}).
$(k_b/2)(2J_b)^2$ is the contribution of in-plane compression of the
lipid bilayer where $k_b$ and $2J_b$ are the compression modulus and
relative area compression of the lipid bilayer.
$G_{sk}=(k_d/2)(2J)^2-\tilde{k}Q$ is the in-plane compression and
shear energy density which comes from the entropic elasticity of the
membrane skeleton. $k_d$ and $\tilde{k}$ are the compression and
shear moduli of the membrane skeleton, respectively. Their values
are experimentally determined as
$k_d=\tilde{k}=4.8\,\mu$N/m.\cite{Lenormand,remktidk} $2J$ and $Q$
are the trace and determinant of the stain tensor of the membrane
skeleton. Because there is no in-plane coupling between the lipid
bilayer and the membrane skeleton in the composite shell model, thus
$J_b$ for the lipid bilayer and $J$ for the membrane skeleton have
no local correlation. In the above subsection, we have mentioned
that the effect of $(k_b/2)(2J_b)^2$ can be replaced with the
surface tension $\lambda=2k_bJ_b$. Considering a closed cell
membrane under osmotic pressure $p$, the free energy can be written
as
\begin{equation}\mathcal{F}=\int G_{cm}\, dA + p\int dV.\label{freeCM}\end{equation}
Similarly to Sec.~\ref{Sec2Dcont}, if we define a displacement
vector $\mathbf{u}$ satisfying Eqs.~(\ref{JQandU1}) and
(\ref{JQandU2}), we can derive the Euler-Lagrange equations
corresponding to the free energy (\ref{freeCM}) as
\begin{eqnarray}&&\hspace{-0.6cm}(\tilde{k}-2k_{d}) \nabla (2J) -\tilde{k}( \diamondsuit ^{2}\mathbf{u}+K\bar\mathbf{u}+\tilde{
\nabla}u_{3})=0,\label{ELcellM1}\\
&&\hspace{-0.6cm}p+2k_{c}[(2H+c_0)(2H^2-c_0H-2K)+2\nabla^{2}H]-2\lambda
H\nonumber\\&&\hspace{-0.34cm}+2H(\tilde{k}-k_{d})(2J)-\tilde{k}\mathfrak{R}:\nabla
\mathbf{u} =0,\label{ELcellM2}\end{eqnarray} where $\bar\mathbf{u}$
and $\diamondsuit ^{2}\mathbf{u}$ are the in-plane components of
$\mathbf{u}$ and $\mathrm{div\,}(\diamondsuit\mathbf{u})$,
respectively. $\mathfrak{R}$ is the curvature tensor related to
Eq.~(\ref{omega13}). $\tilde{ \nabla}$ is called the gradient
operator of the second class, which is shown in our previous
work.\cite{TuJPA04}

Generally speaking, it is difficult to find the analytical solutions
to Eqs.~(\ref{ELcellM1}) and (\ref{ELcellM2}). But we can verify
that a spherical membrane with homogenous in-plane strains satisfy
these equations. The radius $R$ and the homogenous in-plane strain
$\varepsilon$ should obey the following relation:
\begin{equation}\label{spherecem1}
pR^{2}+2(\lambda
+2k_{d}\varepsilon-\tilde{k}\varepsilon)R+{k}_{c}{c}_{0}({c}_{0}R-2)=0.
\end{equation}

\subsubsection{Stability of cell membranes and the function of membrane skeleton}
When the osmotic pressure is beyond some threshold, a closed cell
membrane will lose its stability and change its shape abruptly. The
threshold is called the critical pressure. To obtain it, one should
calculate the second order variation of the free energy
(\ref{freeCM}) in terms of Appendix B. The variational result is
{\small\begin{eqnarray}&&\hspace{-0.3cm}\delta^2 \mathcal{F}=\int
k_{c}[ (\nabla ^{2}\Omega _{3})^{2}+(2H+c_{0})\nabla (2H\Omega
_{3})\cdot \nabla \Omega _{3}] dA\nonumber
\\
&&\hspace{-0.3cm}+\int [4k_{c}(2H^{2}-K)^{2}+k_{c}K(
c_{0}^{2}-4H^{2})+2\lambda
K-2Hp]\Omega _{3}^{2}dA\nonumber \\
&&\hspace{-0.3cm}+\int
[k_{c}(14H^{2}+2c_{0}H-4K-c_{0}^{2}/2)-\lambda ]\Omega _{3}\nabla
^{2}\Omega _{3}dA \nonumber\\
&&\hspace{-0.3cm}-2k_{c}\int (2H+c_{0})[\nabla \Omega _{3}\cdot
\tilde{\nabla}\Omega _{3}+2\Omega _{3}\nabla \cdot
\tilde{\nabla}\Omega
_{3}]dA \nonumber\\
&&\hspace{-0.3cm}-k_{d}\int [ ( \mathbf{v}\cdot \nabla +2H\Omega
_{3})
( \mathrm{div\,}\mathbf{v}-2H\Omega _{3})] dA \nonumber\\
&&\hspace{-0.3cm}+(\tilde{k}/2)\int ( \mathrm{curl\,}\mathbf{v}) ^{2}dA-\tilde{k}\int K\bar\mathbf{v}^{2}dA+\tilde{k}\int \Omega _{3}\tilde{\nabla}%
\cdot \mathbf{v}dA\nonumber\\
&&\hspace{-0.3cm}+\tilde{k}\int 2H\Omega _{3}( \mathrm{div\,}\mathbf{v}%
-2H\Omega _{3}) dA-\tilde{k}\int \Omega _{3}\mathfrak{R}:\nabla
\mathbf{v}dA\label{secondvcellm},\end{eqnarray}}where
$\mathbf{v}=\Omega_1\mathbf{e}_1+\Omega_2\mathbf{e}_2+\Omega_3\mathbf{e}_3$
is the infinitesimal displacement vector of the cell membrane whose
in-plane component is denoted as
$\bar\mathbf{v}=\Omega_1\mathbf{e}_1+\Omega_2\mathbf{e}_2$.

In terms of the Hodge decomposed theorem,\cite{Westenholzbk}
$\mathbf{v}$ can be expressed by two scalar functions $\Omega$ and
$\chi$ as
\begin{equation}\mathbf{v}\cdot d\mathbf{r}=d\Omega +\ast d\chi,\end{equation}
where $\ast$ is the Hodge star.\cite{TuJPA04,Westenholzbk} Then we
have $\mathrm{div\,}\mathbf{v}=\nabla^2\Omega$ and
$\mathrm{curl\,}\mathbf{v}=\nabla^2\chi$. For the spherical cell
membrane satisfying Eq.~(\ref{spherecem1}), Eq.~(\ref{secondvcellm})
can be divided into two parts: one is
\begin{equation}\delta^2 \mathcal{F}_1=(\tilde{k}/2)\int [ (\nabla ^{2}\chi)^2 +\left( 2/R^{2}\right) \chi
\nabla ^{2}\chi ] dA;\end{equation} another is
\begin{eqnarray}
&&\hspace{-0.8cm}\delta^2 \mathcal{F}_2=\int \Omega _{3}^{2}[ 2c_{0}k_{c}/R^{3}+p/R+(4k_{d}-2\tilde{k})/R^{2}] dA \nonumber\\
&&+\int \Omega _{3}\nabla ^{2}\Omega _{3}[
k_{c}c_{0}/R+2k_{c}/R^{2}+pR/2] dA\nonumber \\
&&+\int k_{c}(\nabla ^{2}\Omega
_{3})^{2}dA+[(4k_{d}-2\tilde{k})/R]\int \Omega _{3}\nabla ^{2}\Omega
dA \nonumber \\
&&+k_{d}\int ( \nabla ^{2}\Omega ) ^{2}dA+(\tilde{k}/R^{2})\int
\Omega \nabla ^{2}\Omega dA.
\end{eqnarray}

It is easy to verify that $\delta^2 \mathcal{F}_1$ is always
positive on a spherical surface. Then the stability of the spherical
cell membrane is merely determined by $\delta^2 \mathcal{F}_2$. By
analogy with our previous work,\cite{tupre05} we can prove that
$\delta^2 \mathcal{F}_2$ is also positive if
\begin{equation}p<p_l\equiv \frac{2\tilde{k}( 2k_{d}-\tilde{k}) }{[ k_{d}l(l+1)-
\tilde{k}]
R}+\frac{2k_{c}}{R^{3}}[l(l+1)-c_{0}R],\label{criticalpCM1}\end{equation}
for any integer $l\geq 2$. Thus the critical pressure is
\begin{equation}p_c\equiv\min\{p_l\ (l=2,3,4,\cdots)\}.\label{criticalpCM2}\end{equation}
Obviously, if $\tilde{k}=0$, i.e., the effect of membrane skeleton
vanishes in the cell membrane, $p_c$ degenerates into the critical
pressure (\ref{criticalpsc}) of a spherical lipid vesicle.

When $\tilde{k}k_{d}( 2k_{d}-\tilde{k}) R^{2}/k_{c}( 6k_{d}-\tilde{
k}) ^{2}>1$, the critical pressure is derived from
Eqs.~(\ref{criticalpCM1}) and (\ref{criticalpCM2}) as
\begin{equation}p_{c}=(4/R^{2})\sqrt{(\tilde{k}/k_{d})( 2k_{d}-\tilde{k})
k_{c}}\label{criticalpCM3}\, .\end{equation} As an example, let us
consider a cell membrane with typical values of
$\tilde{k}=k_{d}=4.8\,\mu$N/m,\cite{Lenormand} $k_{c}=10^{-19}$ J,
and $R \approx 10\,\mu m$. Through a simply manipulation, we find
that $\tilde{k}k_{d}( 2k_{d}-\tilde{k}) R^{2}/k_{c}( 6k_{d}-\tilde{
k}) ^{2}\gg 1$, and so Eq.~(\ref{criticalpCM3}) holds, from which we
obtain the critical pressure $p_c=0.03$ Pa. However, if the membrane
skeleton vanishes, $\tilde{k}=0$, we calculate $p_c=0.001$ Pa from
Eqs.~(\ref{criticalpCM1}) and (\ref{criticalpCM2}). This example
reveals a mechanical function of membrane skeleton: it highly
enhances the stability of cell membranes.

As a byproduct, Eq.~(\ref{criticalpCM3}) also gives the critical
pressure
\begin{equation}p_c={\sqrt{4/3(1-\nu ^{2})}}\,\  Y(h/R)^{2}\end{equation}
for a spherical thin solid shell of 3D isotropic materials if we
take $k_c$, $k_d$, and $\tilde{k}$ as
Eqs.~(\ref{ratiokbkc0})--(\ref{ratiokbkc}). This formula is the same
as the classic strict result obtained by Pogorelov from the other
method.\cite{Pogorelovbook}

\section{Application of Elastic theory in nano-structures\label{SecApNanoStuct}}
In the last section, we have expatiated on the application of
Elastic theory in bio-structures. In this section, we will discuss
whether and to what extent this theory can be applied to
nano-structures, especially the graphitic structures, such as
graphene and carbon nanotubes.

\begin{figure}[pth!]
\includegraphics[width=7.5cm]{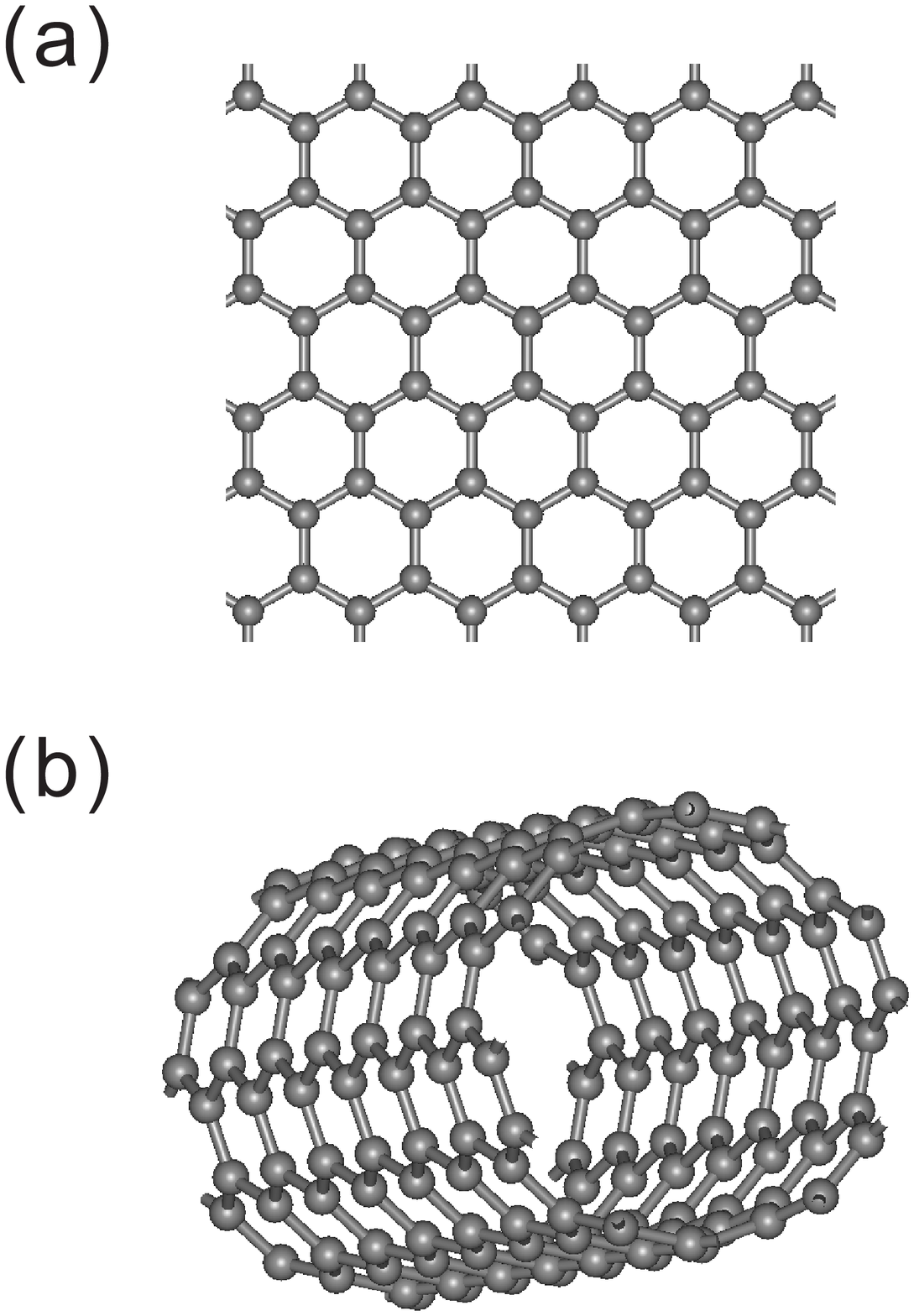}\caption{\label{figgraphene}
(a) Graphene. (b) Single-walled carbon nanotube.}
\end{figure}

\subsection{Graphene}
Graphene is a single layer of carbon atoms with a 2D honeycomb
lattice as shown in Fig.~\ref{figgraphene}a. It has been a rapidly
rising star in the material science and condensed-matter
physics\cite{GeimNM07} since it was successfully cleaved from buck
graphite.\cite{NovoselovSci04} It is found that the free-standing
graphene might be a strictly 2D atomic crystal which is stable under
ambient conditions.\cite{NovoselovPNAS05} However, Mermin has
theoretically proved that the 2D crystalline order could not exist
at finite temperature.\cite{Mermin68} There are two possible ways to
solve this paradox: (i) The graphene might not be a perfect 2D
crystal. Recently, Meyer \textit{et al.} have investigated the
elaborate structure of suspended graphene sheets and found that the
graphene sheets are not genuine flat.\cite{NovoselovN07} They also
argue that the graphene sheets could be stabilized by the
out-of-plane deformation in the third dimension resulting from the
the thermal fluctuations.\cite{NovoselovN07} Fasolino \textit{et
al.} have also addressed the height fluctuations by means of Monte
Carlo simulations.\cite{Fasolino07} Their result at room temperature
is in good agreement with the experiment mentioned above. (ii)
Mermin theorem is valid for power-law potentials of the
Lennard-Jones type while the interaction between nearest neighbor
atoms (covalent bond) in the graphene might not be of this
type.\cite{Garcia07}

To fully understand the experimental result and possible stable
mechanism in theory, we will address the Lenosky lattice
model\cite{LenoskyN92} and its revised form as follows.

\subsubsection{Revised Lenosky lattice model and its continuum limit}
We start from the concise formula proposed by Lenosky \textit{et
al.} in 1992 to describe the deformation energy of a single layer of
curved graphite\cite{LenoskyN92}
\begin{eqnarray} \label{engleno}
&&\hspace{-1.1cm}E_g=\frac{\epsilon_{0}}{2}
\sum_{(ij)}(r_{ij}-r_{0})
^2+\epsilon_{1} \sum_{i}(\sum_{(j)}\mathbf{u}_{ij})^2 \nonumber\\
&&\hspace{-0.55cm}+\epsilon_{2} \sum_{(ij)}(1-\mathbf{n}_{i} \cdot
\mathbf{n}_{j})+\epsilon_{3} \sum_{(ij)}(\mathbf{n}_{i} \cdot
\mathbf{u}_{ij})(\mathbf{n}_{j} \cdot
\mathbf{u}_{ji}).\end{eqnarray} The first two terms are the
contributions of bond length and bond angle changes to the energy.
The last two terms are the contributions from the $\pi$-electron
resonance. In the first term, $r_{0}$ is the initial bond length of
planar graphite, and $r_{ij}$ is the bond length between atoms $i$
and $j$ after the deformations. In the remaining terms,
$\mathbf{u}_{ij}$ is a unit vector pointing from atom $i$ to its
neighbor $j$, and $\mathbf{n}_{i}$ is the unit vector normal to the
plane determined by the three neighbors of atom $i$. The summation
${\sum_{(j)}}$ is taken over the three nearest neighbor atoms $j$ to
atom $i$, and ${\sum_{(ij)}}$ taken over all the nearest neighbor
atoms. The parameters
$(\epsilon_{1},\epsilon_{2},\epsilon_{3})=(0.96,1.29,0.05)$~eV were
determined by Lenosky \textit{et al.} \cite{LenoskyN92} through
local density approximation. The value of $\epsilon_{0}$ was given
by Zhou \textit{et al.} as $\epsilon_{0}=57\,\,{\rm eV/\AA^2}$
through the force-constant method.\cite{zhouxPB01}

In the above energy form, the second term requires that the energy
cost due to in-plane bond angle changes is the same as that due to
out-of-plane bond angle changes. However, the experiment by
inelastic neutron scattering techniques reveals that the energy
costs due to in-plane and out-of plane bond angle changes are quite
different from each other.\cite{nicklowPRB72} To describe this
effect, we revise the Lenosky lattice model as
\begin{eqnarray}
&&\hspace{-0.5cm}E_g=\frac{\epsilon
_{0}}{2}\sum_{(ij)}(r_{ij}-r_{0})
^{2}+\epsilon _{1t}\sum_{i}\sum_{(j<k)}( \mathbf{u}_{ij}^{t}\cdot \mathbf{u}%
_{ik}^{t}+{1}/{2}) ^{2}\nonumber\\&&+\epsilon
_{1n}\sum_{i}(\sum_{(j)}\mathbf{u}_{ij}^{n})^{2}+\epsilon
_{2}\sum_{(ij)}( 1-\mathbf{n}_{i}\cdot
\mathbf{n}_{j}),\label{revisengleno}\end{eqnarray} where
$\mathbf{u}_{ij}^{t}=\mathbf{u}_{ij}-(\mathbf{n}_{i}\cdot
\mathbf{u}_{ij})\mathbf{n}_{i}$ and $\mathbf{u}_{ij}^{n}
=\mathbf{n}_{i}\cdot \mathbf{u}_{ij}$. If the three nearest neighbor
atoms to atom $i$ are labeled as 1,2,3, the summation $\sum_{(j<k)}$
is understood as $\sum_{1\leq j<k\leq 3}$. The second and third
terms of Eq.~(\ref{revisengleno}) represent the energy costs due to
in-plane and out-of-plane bond angle changes, respectively. We have
omitted the term $\epsilon_{3} \sum_{(ij)}(\mathbf{n}_{i} \cdot
\mathbf{u}_{ij})(\mathbf{n}_{j} \cdot \mathbf{u}_{ji})$ relative to
the original Lenosky model (\ref{engleno}), because its contribution
is very small in terms of the results by Lenosky \textit{et al.}.

The parameters in Eq.~(\ref{revisengleno}) are determined by fitting
the total energy of variously perturbed configurations of
$\sqrt{7}\times \sqrt{7}$ unit cell of graphite (14 atoms). The
total energy is obtained through the first-principles calculations
(the ABINIT package \cite{Gonzexcms}). The calculations are carried
by taking Troullier-Martins pseudopotentials, \cite{Troullier}
plane-wave energy cutoff of 50 Hartree, and $4\times 4\times 1$
Monkhorst-Pack k-points \cite{MonkhorstPack} in Brillouin-zone. The
exchange-correlation energy are treated within the local-density
approximation in the Ceperley-Alder form \cite{CeperleyAlder} with
the Perdew-Wang parametrization. \cite{PerdewWang} Our result is
$r_0=1.41$~\AA, $\epsilon_{0}=46.34$~eV/\AA$^2$,
$\epsilon_{1t}=4.48$~eV, $\epsilon_{1n}=1.04$~eV, and
$\epsilon_{2}=1.24$~eV. The value of $\epsilon_{0}$ is a little
smaller than that obtained by Zhou \textit{et al.}\cite{zhouxPB01}
from force constant method. The values of $\epsilon_{1n}$ and
$\epsilon_{2}$ are very close to those of $\epsilon_1$ and
$\epsilon_{2}$ obtained by Lenosky \textit{et al.}\cite{LenoskyN92}
from local density approximation. The key reason is that the main
energy contribution in the configurations discussed by Lenosky
\textit{et al.} comes from the third and fourth term in
Eq.~(\ref{revisengleno}).

Now let us derive the continuum limit form of the revised Lenosky
lattice model (\ref{revisengleno}) by analogy with the method in our
previous work.\cite{OuYangPRL97,TuzcPRB02} Now consider a curved
graphene and take a fictitious smooth surface such that all carbon
atoms are on that surface. The in-plane stain can be expressed as
$\mathfrak{E}_i=\left(\begin{array}{cc} \varepsilon_{11} & \varepsilon_{12} \\
\varepsilon_{12} & \varepsilon_{22} \end{array} \right)$ in the
local frame $\{\mathbf{e}_1,\mathbf{e}_2,\mathbf{e}_3\}$ at atom
$i$. The bond vector $\mathbf{r}_{ij}$ from atom $i$ to its neighbor
$j$ after the deformations and the initial bond vector
$\mathbf{r}_{ij}^{0}$ before the deformations satisfy
$\mathbf{r}_{ij}=(\mathbf{I}+\mathfrak{E}_i)\cdot\mathbf{r}_{ij}^{0}$,
where $\mathbf{I}$ is the unit matrix. The initial bond vectors
$\mathbf{r}_{ij}^{0}$ can be expanded to the order of
$O(r_0^2\kappa^2)$ as \cite{Carmobook}
\begin{eqnarray}
&&\mathbf{r}_{ij}^0=(1-r_0^2\kappa^2_j/6)r_0\mathbf{T}_j+(\kappa_j
\tau _j r_0^3 /6)\mathbf{B}_j  \nonumber \\
&&\hspace{0.5cm}+[r_0\kappa_j/2+(r_0^2/6)d\kappa_j/ds]r_0\mathbf{N}_j,
\label{bondcv2nd}\end{eqnarray} where $j$=1, 2, 3 denote three
$sp^2$-bond curves from atom $i$ to one of its three neighbor atoms
$j$ on the graphene surface. The symbols $\mathbf{T}_j$,
$\mathbf{N}_j$, and $\mathbf{B}_j$ represent the unit tangential,
normal, and binormal vectors of the bond curve from $i$-atom to
$j$-atom, which satisfy the Frenent theorem Eq.~(\ref{framefrenet}).
$\kappa$, $\tau$ refer to the curvature and torsion while $s$ is the
arc-length parameter along the bond curve. Assume the $sp^2$-bond
along the geodesic curve of the graphene surface. The vectors
$\mathbf{T}_j$ and $\mathbf{B}_j$ can be expressed by
$\mathbf{T}_j=\cos \theta_j\mathbf{e}_1+\sin \theta_j\mathbf{e}_2$
and $\mathbf{B}_j=-\sin \theta_j\mathbf{e}_1+\cos
\theta_j\mathbf{e}_2$, where $\theta_j$ is the rotating angle from
$\mathbf{e}_1$ to $\mathbf{T}_j$. We have the expressions of ${\bf
u}_{ij}=\mathbf{r}_{ij}/r_{ij}$ and $\mathbf{n}_i=\mathbf{N}_j$ with
$r_{ij}=|\mathbf{r}_{ij}|$ for the deformed graphene. Then
Eq.~(\ref{revisengleno}) is transformed into the continuum limit up
to the second-order magnitudes of $\varepsilon_{11}$,
$\varepsilon_{22}$, $\varepsilon_{12}$ and $r_0\kappa$ as
\begin{equation}E_g=\int \left[\frac{k_c}{2}(2H)^2-\bar{k}K+\frac{k_d}{2}(2J)^2-\tilde{k}Q\right] dA,\label{freeng-gr}\end{equation}
with four parameters
\begin{eqnarray} \label{kcvalue}
&&k_{c}=(9\epsilon _{1n}+6\epsilon _{2}) r_{0}^{2}/8\Omega _{0},\\
\label{kbarvalue}
&&\bar{k}={3\epsilon _{2}r_{0}^{2}}/{4\Omega_{0}},\\
\label{kdvalue}&&k_{d}=9(\epsilon _{0}r_{0}^{2}+3\epsilon _{1t})/
16\Omega _{0},\\
\label{ktildvalue}&&\tilde{k}=3( \epsilon _{0}r_{0}^{2}+9\epsilon
_{1t})/8\Omega _{0},
\end{eqnarray}
where $\Omega_0=3\sqrt{3}r_0^2/4$ is the occupied area per atom. The
continuum form (\ref{freeng-gr}) has first derived in our previous
work \onlinecite{TuzcPRB02} which is, in fact, the natural
conclusion of the symmetry of graphene:\cite{TuJCTN06} The curved
graphene comprises a lot of hexagons which has approximately local
hexagonal symmetry. In fact, 2D structures with hexagonal symmetry
are 2D isotropic.\cite{nyebook} Thus the elasticity of the graphene
can be reasonably described by the shell theory of 2D isotropic
materials mentioned in Sec.~\ref{Sec2Dcont} and so its energy has
the form of Eq.~(\ref{freeng-gr}). We also notice that a flaw in the
coefficient before $\epsilon _{1}$ in the expression of $k_c$ in our
previous work \onlinecite{TuzcPRB02}.

Using the values of $r_0$, $\epsilon _{1t}$, $\epsilon _{1n}$, and
$\epsilon _2$ obtained from the first-principles calculations, we
have $k_c=1.62$~eV, $\bar{k}=0.72$~eV, $k_d=22.97$~eV/\AA$^2$, and
$\tilde{k}=19.19$~eV/\AA$^2$. Because the results of
first-principles calculation are applicable for zero temperature,
only the results derived from the experiments at low temperature can
be used as reference values to compared with them. The value of
$k_c$ is close to the value 1.77~eV estimated by Komatsu
\cite{KomatsuJPSJ55,KomatsuJPCS58,NihiraPRB03} at low temperature
(less than 60~K). The value $\tilde{k}/k_d=0.83$ is quite close to
the experimental value 0.8 derived from the in-plane elastic
constants of graphite. \cite{BlakesleeJAP70} The elastic properties
of graphene can be described by Eq.~(\ref{freeng-gr}) with four
parameters $k_c$, $\bar{k}$, $k_d$, and $\tilde{k}$, where the
energy density is the same as Eq.~(\ref{edstyshell}), the free
energy density of solid shell with 2D isotropic materials. Since
$\bar{k}/k_c=0.44$ is much smaller than $\tilde{k}/k_d=0.83$, the
graphene cannot be regarded as a solid shell with 3D isotropic
materials as Ref.~\onlinecite{TuzcPRB02}.

\subsubsection{Intrinsic roughening in graphene at temperature T}
Let us consider the freely suspended graphene which is almost a flat
layer with the area $L^2$. The small out-of-plane displacement is
denoted by $w$. The energy (\ref{freeng-gr}) is transformed into
\begin{equation}E_g= (k_c/2)\int (\nabla^2w)^2\, d^2\mathbf{x},\label{engflatg}\end{equation}
where $\mathbf{x}\equiv(x_1,x_2)$ represents the point on the
graphene plane before deformations.

Adopting the Fourier series
\begin{equation}w(\mathbf{x})=(1/L)\sum_{\mathbf{q}}\tilde{w}_\mathbf{q}\exp\{
i\mathbf{q}\cdot\mathbf{x}\},\end{equation} with
$\mathbf{q}\equiv(2l\pi/L,2n\pi/L)$, we transform
Eq.~(\ref{engflatg}) into
\begin{equation}E_g=(k_c/2)\sum_{\mathbf{q}}\mathbf{q}^4|\tilde{w}_\mathbf{q}|^2,\end{equation}
and then the corresponding partition function is derived as
\begin{equation}\mathcal{Z}=\int \prod_\mathbf{q}d\tilde{w}_\mathbf{q}\exp(-E_g/T)=\prod_\mathbf{q}\sqrt{2\pi T/k_c\mathbf{q}^4},\end{equation}
where the Boltzmann constant has been set to 1. It follows that the
equipartition theorem:
\begin{equation}\langle(k_c/2)\mathbf{q}^4|\tilde{w}_\mathbf{q}|^2\rangle=-T\partial\ln\mathcal{Z}/\partial\ln \mathbf{q}^4=T/2,\end{equation}
where $\langle .\rangle$ represents the ensemble average. The above
equation is equivalent to
\begin{equation}\langle|\tilde{w}_\mathbf{q}|^2\rangle=T/k_c\mathbf{q}^4.\end{equation}
Similarly, $\langle w^2 \rangle$ is derived as
\begin{equation}\langle w^2
\rangle=\sum_{\mathbf{q}}\frac{\langle|\tilde{w}_\mathbf{q}|^2\rangle}{L^2}=\frac{TL^2}{16\pi^4k_c}\sum_{ln}\frac{1}{(l^2+n^2)^2}.
\end{equation}
Through simply numerical manipulations, we have\cite{NetoKim}
\begin{equation}\langle w^2\rangle\simeq\frac{TL^2}{150k_c},\label{msqhightg}\end{equation}
for the graphene contains more than 100 atoms.

In terms of Ref.~\onlinecite{NihiraPRB03}, we estimate $k_c\approx
0.46$~eV at $T=300$~K. Substituting it into Eq.~(\ref{msqhightg})
and taking $L=25$~nm as the experiment \onlinecite{NovoselovN07}, we
have $\sqrt{\langle w^2\rangle}\approx 0.5$~nm. This value is a
little smaller than the largest out-of-plane deformation $1$~nm in
the experiment. However, they are consistent with each other because
0.5~nm is the mean square value which should be smaller than the
largest out-of-plane deformation in the experiment.

\subsection{Carbon nanotube}
There are two kinds of carbon nanotubes: single- and multi- walled
carbon nanotubes, which are synthesized in the last decade of 20
century.\cite{Iijima1,Iijima93} Simply speaking, a single-walled
carbon nanotube (SWNT) can be regarded as a seamless cylinder
wrapped up from a graphitic sheet, as shown in
Fig.~\ref{figgraphene}b, whose diameter is in nanometer scale and
length from tens of nanometers to several micrometers if we ignore
its two end caps. A multi-walled carbon nanotube (MWNT) consists of
a series of coaxial SWNTs with layer distance about 3.4~\AA.

SWNTs can be expressed as a pair of integers (n,m), so called index,
in terms of the wrapping rule. They are divided into two classes:
achiral tubes if $n=m$ or $nm=0$ and chiral tubes for others.
\cite{Saitobook} The electronic properties of SWNTs depend
sensitively on the index: \cite{MintmirePRL} they are metallic if
$n-m$ is multiple of 3, else semiconductor. SWNTs also possess many
novel mechanical properties,\cite{KrishnanPRB} in particular high
stiffness and axial strength, which are not sensitive to the tube
diameters and chirality. MWNTs have the similar mechanical
properties to SWNTs. \cite{TreacyN96,WongSci97} In this section, we
will review the theoretical and numerical results on the elastic
properties of carbon nanotubes, and then discuss how the
low-dimensional elastic theory mentioned in Sec.~\ref{ETLDC} can be
applied in carbon nanotubes.

\subsubsection{General review on the elasticity of carbon nanotubes}
The early researches on the elasticity of carbon nanotubes are
focused on their Young's modulus $Y$ and Poisson ratio $\nu$. A SWNT
is a single layer of carbon atoms. What is the thickness $h$ of the
atomic layer? It is a widely controversial question. Three typical
values of the thickness listed in Table~\ref{Youngmod} are adopted
or obtained in the previous literature Refs.~\onlinecite{Yakobson},
\onlinecite{Lujp}, \onlinecite{TuzcPRB02} and
\onlinecite{KudinPRB01}--\onlinecite{ZhouGCPL01}. The first one is
about 0.7~\AA\ obtained from fitting the atomic scale model with the
elastic shell theory of 3D isotropic
materials.\cite{KudinPRB01,ZhouxPRB2000,ZhangLCPRB03,Pantano,PantanoJMP,ChenGaoN06,ZhengQPRL05}
The second one is about 1.4~\AA\ derived from molecular dynamics or
finite element method.\cite{SearsPRB04,TserpesCPB05} The third one
is about 3.4~\AA\ adopting the layer distance of bulk
graphite.\cite{HernandezAPA,ShenPRB05,LiChouIJSS03,WenXingPB04,ZhouGCPL01}
Recently, Huang \textit{et al.} have investigated the effective
thickness of SWNTs and found it depends on the type of
loadings.\cite{HuangPRB06}

\begin{table}[!htp]
\caption{Young's modulus $Y$ (unit in TPa), Poisson ratio $\nu$ and
effective thickness $h$ (unit in \AA). (MD = molecular dynamics; TB
= tight-binding; SM = structure mechanics; FEM = finite element
method; LDA = local density approach)\label{Youngmod}}
\begin{ruledtabular}
\begin{tabular}{cccccc}
Authors & $Y$ & $\nu$ & $h$ & Method & Refs.\\
\hline Yakobson \textit{et al.}& 5.5 & 0.19 & 0.66 & MD &
\onlinecite{Yakobson}\\
Tu \& Ou-Yang & 4.7& 0.34&0.75& LDA & \onlinecite{TuzcPRB02}\\
Kudin \textit{et al.}& 3.9 & 0.15 & 0.89 & \textit{ab initio}&
\onlinecite{KudinPRB01}\\
Zhou \textit{et al.}& 5.1 & 0.24 & 0.74 & TB
&\onlinecite{ZhouxPRB2000}\\
Vodenitcharova \textit{et al.}& 4.9 &--& 0.62 & ring theory &
\onlinecite{ZhangLCPRB03}\\
Pantano \textit{et al.}&4.8&0.19&0.75&SM \& FEM &\onlinecite{Pantano,PantanoJMP}\\
Chen and Cao & 6.8 &--&0.80 & SM & \onlinecite{ChenGaoN06}\\
Wang \textit{et al.}&5.1&0.16&0.67&\textit{ab initio}&\onlinecite{ZhengQPRL05} \\
Sears \& Batra & 2.5 & 0.21 &1.34 & MD & \onlinecite{SearsPRB04}\\
Tserpes \textit{et al.} &2.4&--&1.47&FEM&\onlinecite{TserpesCPB05}\\
Lu & 1.0& 0.28 &3.4& MD & \onlinecite{Lujp}\\
Hernandez \textit{et al.}&1.2&0.18 &3.4&TB & \onlinecite{HernandezAPA}\\
Shen \& Li & 1.1 & 0.16 &3.4 & force-field & \onlinecite{ShenPRB05}\\
Li \& Chou & 1.0 & -- &3.4& SM &
\onlinecite{LiChouIJSS03}\\
Bao \textit{et al.}&0.9&--&3.4&MD &
\onlinecite{WenXingPB04}\\
Zhou \textit{et al.}&0.8&0.32 &3.4&LDA&\onlinecite{ZhouGCPL01}
\end{tabular}
\end{ruledtabular}
\end{table}

The size- or chirality-dependent elastic properties of SWNTs have
also been discussed by molecular mechanics model
\cite{ChangJMPS03,LiChouPRB04,ChangAPL05} and \textit{ab initio}
calculations.\cite{ZhengQPRL05,MoriJJAP05} The common conclusion is
that the Young's modulus and Poisson ratio depend weakly on the
diameter and chirality of SWNTs if the diameter is larger than 1~nm.
Only for very small SWNTs, the size and chirality effect is evident.
The SWNTs synthesized in the laboratory have usually the diameters
larger than 1~nm; thus the size and chirality effect can be
neglected safely.

The axial tension properties of MWNTs depend on the layer number of
MWNTs for the small layer number and approach quickly to the
properties similar to the bulk graphite.
\cite{TuzcPRB02,GovindjeeSSC99,LiuJPD04}

The buckling and stability of carbon nanotubes under pressure or
bending is a hot topic in the recent researches, where the critical
pressure, moment or the equivalent quantity, critical strain, are
highly concerned. A long enough carbon nanotube under an axial
loading might be regarded as a Euler rod and the axially critical
strain is \cite{landau}
\begin{equation}\varepsilon_{zc}^{rod}=\alpha\pi^2 \mathcal{I}/AL^2\propto (\rho/L)^2,\label{epsceulerax}\end{equation}
where $L$, $\rho$ and $A$ are the length, radius and cross-sectional
area of the carbon nanotube, respectively. $\mathcal{I}$ is the
moment of inertia of the nanotube. The value of $\alpha$ depends on
the boundary conditions of the carbon nanotube. This relation has
been investigated by atomic-scale finite element method
\cite{LiuCMAPE04,GuoXJAM07,WangCMJAP06} and molecular dynamics
method or \textit{ab initio} calculations.
\cite{HarikCMS02,LiewPRB04,SearsPRB06,CaoChenN06,VaradanSMS05} The
basic numerical result is that the tube exhibits rod-like buckling
behavior as the right-handed side of Eq.~(\ref{epsceulerax}) if
$L\gg \rho$. The Timoshenko beam theory, a more complicated theory
than Euler rod theory, is also employed to discuss the buckling of
MWNTs.\cite{ZhangJEM06} The difference between the results of both
theories vanishes for large value of $L/\rho$.

For a short carbon nanotube under axial loading, the continuous
shell model of 3D isotropic materials are widely
used.\cite{Yakobson,RuPRB2000,LiuCMAPE04,GuoXJAM07,WangCMJAP06,LiewPRB04,SearsPRB06,CaoChenN06,XiaoJAP04,DasCMS02}
The axially critical strain of a short SWNT is \cite{Pogorelovbook}
\begin{equation}\varepsilon_{zc}^{shell}=[\alpha/\sqrt{3(1-\nu^2)}] (h/\rho)\propto \rho^{-1},\label{epscshellax}\end{equation}
where $\rho$ and $h$ are the radius and effective thickness of the
SWNT, respectively. $\nu$ is the Poisson ratio of the SWNT. The
value of $\alpha$ depends on the boundary condition of the carbon
nanotube. For a short MWNT, the above relation is applicable for the
outmost layer of the tube because the inter-layer interaction of
MWNTs is very small.\cite{HeLiewJMPS05} It has also been
investigated by atomic-scale finite element method,
\cite{LiuCMAPE04,GuoXJAM07,WangCMJAP06,DasCMS02} molecular dynamics
method,\cite{Yakobson,LiewPRB04,SearsPRB06,CaoChenN06,XiaoJAP04} and
nanoindent experiment.\cite{WatersAPL04,WatersCST06} It is found
that the tube displays indeed the shell-like buckling behavior as
the right-handed side of Eq.~(\ref{epscshellax}) for the tube aspect
ratio $L/\rho<10$.

The stability of a long SWNT under radial hydrostatic pressure might
also be described by the continuous shell model of 3D isotropic
materials, and the critical pressure is \cite{Pogorelovbook}
\begin{equation}p_{cr}^{shell}\propto \rho^{-3},\label{pcrshellrd}\end{equation}
where $\rho$ is the radius of the SWNT. This relation has recently
been confirmed by Hasegawa and Nishidate \cite{HasegawaPRB06}
through \textit{ab initio} calculations. The stability of a MWNT
under radial hydrostatic pressure might also has the similar
relation as Eq.~(\ref{pcrshellrd}) if only we take $\rho$ as the
outmost radius of the MWNT, because the transverse elasticity of
MWNTs \cite{PalaciPRL05,DaiEPJB06} is much weaker than the in-plane
elasticity of the outmost single layer of tube.

Bending can also result in the buckling of SWNTs. The kink
phenomenon in a SWNT under pure bending has been investigated
through molecular dynamics simulations and finite element
method.\cite{Yakobson,CaoChenPRB06} The critical curvature can be
described as
\begin{equation}\kappa_{cr}=\varepsilon_{zc}^{shell}/\rho\propto \rho^{-2},\label{kappacrbd}\end{equation}
where $\rho$ is the radius of the SWNT. The kink phenomenon in a
MWNT under pure bending satisfies the similar relation to
Eq.~(\ref{kappacrbd}) with small correction due to inter-layer van
de Waals interactions \cite{ChangGPRB05,WangHCJPRB05,WangYangPRB06}
if only we take $\rho$ as the outmost radius of the MWNT.

Here we would not further discuss the problems on the buckling of
MWNTs embedded in an elastic
medium,\cite{RuCQJMPS01,KitipornchaiJAP05,ZhangLiIJMS06,YangMSMSE06,WangXIJSS07,HanEJMA03}
the postbuckling behavior and the plastic properties of carbon
nanotubes,\cite{ShenIJSS04,ShengZPRB06,LeungJAP06,YaoHanEJMA07,WangXYCPB04,ZhangLammertPRL98,SrivastavaPRL99}
as well as the mechanical properties of nanotube
composites,\cite{LauJCTN04,LauCPB04,LauCPB06,LustiMS04,VodenitIJSS06,VargheseMAMS06}
rather than recommend gentle readers to consult the corresponding
literature.

\subsubsection{What are the fundamental quantities for SWNTs?}
As mentioned above, different thickness leads to different Young's
modulus (see Table~\ref{Youngmod}), which implies that the Young's
modulus and thickness of SWNTs are not well-defined physical
quantities.\cite{RajendranJCTN06} However, the in-plane Young's
modulus $Y_s=Yh$ has the similar value 22~eV/\AA$^2$. Thus it is a
more well-defined quantity than the Young's modulus and the
thickness. Here we may ask: what are the fundamental quantities for
SWNTs?

A SWNT is also a single layer of graphite, whose deformation energy
can be also described as the revised Lenosky model
(\ref{revisengleno}). The corresponding continuum limit is
Eq.~(\ref{freeng-gr}) which contains four elastic constants $k_c$,
$\bar{k}$, $k_d$, and $\tilde{k}$. These four quantities avoid the
controversial thickness of SWNTs. We suggest to use them as the
fundamental quantities for SWNTs from which we can obtain some
reduced quantities as follows.

Let us consider a cylinder under an axial loading with line density
$f$ along the circumference. The corresponding axial and
circumferential strains are denoted as $\varepsilon_{11}$ and
$\varepsilon_{22}$. With Eq.~(\ref{freeng-gr}), the free energy of
this system is written as
\begin{equation}\mathcal{F}\approx 2\pi \rho L[(k_d/2)(\varepsilon_{11}+\varepsilon_{22})^2
-\tilde{k}\varepsilon_{11}\varepsilon_{22}-
f\varepsilon_{11}]\end{equation} where $L$ and $\rho$ are the length
and radius of the SWNT. The in-plane Young's modulus and Poisson
ratio can be defined as $Y_s=f/\varepsilon_{11}$ and
$\nu_s=-\varepsilon_{22}/\varepsilon_{11}$. From
$\partial\mathcal{F}/\varepsilon_{11}=0$ and
$\partial\mathcal{F}/\varepsilon_{22}=0$, we derive
\begin{eqnarray}
&&Y_s=\tilde{k}(2-\tilde{k}/k_d)=22.35\ \mathrm{eV/\AA}^2,\label{Youngsmodip}\\
&&\nu_s=1-\tilde{k}/k_d=0.165,\label{ippoisratio}
\end{eqnarray}
where the value of $Y_s$ is close to the in-plane Young's modulus
derived from Table~\ref{Youngmod}. It is in between
20--23~eV/\AA$^2$ obtained by S\'{a}nchez-Portal \textit{et
al.}.\cite{RubioPRB99} It is much larger than the value
15~eV/\AA$^2$ obtained by Arroyo \textit{et al.}\cite{ArroyoPRB04}
and Zhang \textit{et al.},\cite{ZhangIJSS02} and 17~eV/\AA$^2$ by
Caillerie \textit{et al.},\cite{CaillerieJE06} but smaller than
34.6~eV/\AA$^2$ for armchair tube by Wang.\cite{WangIJSS04} The
value of $\nu_s$ is close to the value 0.16--0.19 obtained by
Yakobson \textit{et al.},\cite{Yakobson} Kudin \textit{et
al.},\cite{KudinPRB01} Pantano \textit{et
al.},\cite{Pantano,PantanoJMP} Wang \textit{et
al.},\cite{ZhengQPRL05} Hernandez \textit{et
al.},\cite{HernandezAPA} and Shen \textit{et al.}.\cite{ShenPRB05}

The other quantity, the bending rigidity $D$, is also widely
discussed in literature. In terms of Eq.~(\ref{freeng-gr}), the
energy per area of a SWNT without the in-plane strains can be
expressed as
\begin{equation}G_g=k_c/2\rho^2\equiv D/2\rho^2.\end{equation}
Thus the bending rigidity
\begin{equation}D= k_c=1.62\ \mathrm{eV},\label{outbendrig}\end{equation}
which is quite close to the value 1.49--1.72~eV obtained by Kudin
\textit{et al.}\cite{KudinPRB01} and S\'{a}nchez-Portal \textit{et
al.}\cite{RubioPRB99} through \textit{ab initio} calculations. It is
a little larger than the values 0.85--1.22~eV obtained Yakobson
\textit{et al.},\cite{Yakobson} Pantano \textit{et
al.},\cite{Pantano,PantanoJMP} and Wang. \cite{WangIJSS04}

In terms of Eqs.~(\ref{Youngsmodip})--(\ref{outbendrig}), we can
infer the values of $k_d$, $\tilde{k}$, $k_c$ from the previous
literature, which are listed in Table~\ref{elasticpar}. There is
still lack of literature on $\bar{k}$ except our previous
work\cite{OuYangPRL97,TuzcPRB02,TuJCTN06} and the present review.
More work on $\bar{k}$ would be highly appreciated in the future.

\begin{table*}[!htp]
\caption{The values of $Y_s$, $\nu_s$, $k_d$, $\tilde{k}$, $k_c$ and
$\bar{k}$. (MD = molecular dynamics; TB = tight-binding; SM =
structure mechanics; FEM = finite element method; LDA = local
density approach; CTIP=continuum theory of interatomic
potential)\label{elasticpar}}
\begin{ruledtabular}
\begin{tabular}{ccccccccc}
Authors & $Y_s$ (eV/\AA$^2$) & $\nu_s$ & $k_d$ (eV/\AA$^2$)& $\tilde{k}$ (eV/\AA$^2$) &$k_c$ (eV) & $\bar{k}$ (eV) & Method & Refs.\\
\hline Yakobson \textit{et al.}& 22.69 & 0.19 & 23.54 &19.06
&0.85&-- & MD &
\onlinecite{Yakobson}\\
Tu \& Ou-Yang & 22.03 & 0.34 &24.88 &16.44 & 1.17&0.75& LDA & \onlinecite{TuzcPRB02}\\
Tu \& Ou-Yang &21.63 & 0.18 &22.35& 18.33 & 1.30 & 0.88&LDA&\onlinecite{TuJCTN06}\\
Kudin \textit{et al.}& 21.69 & 0.15 & 22.19 &18.86& 1.49--1.53 & --&
\textit{ab initio}&
\onlinecite{KudinPRB01}\\
Zhou \textit{et al.}& 23.59 & 0.24 & 25.03&19.02 &1.14&-- & TB
&\onlinecite{ZhouxPRB2000}\\
Pantano \textit{et al.}&22.5&0.19&23.34 &18.91 &1.09&--&SM \& FEM &\onlinecite{Pantano,PantanoJMP}\\
Chen and Cao &34.38 &--&--&--&--&--& SM & \onlinecite{ChenGaoN06}\\
Wang \textit{et al.}&21.36&0.16&21.92&18.41&0.82&--&\textit{ab initio}&\onlinecite{ZhengQPRL05} \\
Sears \& Batra & 20.94 & 0.21 &21.90&17.30&3.28&--& MD & \onlinecite{SearsPRB04}\\
Tserpes \textit{et al.} &22.05&--&--&--&--&--&FEM&\onlinecite{TserpesCPB05}\\
Lu & 21.25& 0.28 &23.06&16.60&--&--& MD & \onlinecite{Lujp}\\
Hernandez \textit{et al.}&25.50&0.18 &26.35&21.61&--&--&TB & \onlinecite{HernandezAPA}\\
Shen \& Li & 23.38 & 0.16 &23.99&20.16&--&-- & force-field & \onlinecite{ShenPRB05}\\
Li \& Chou & 21.25 & --&--&--&--&--& SM &
\onlinecite{LiChouIJSS03}\\
Bao \textit{et al.}&19.13&--&--&--&--&--&MD &
\onlinecite{WenXingPB04}\\
Zhou \textit{et
al.}&17.00&0.32&18.94&12.88&--&--&LDA&\onlinecite{ZhouGCPL01}\\
S\'{a}nchez-Portal&19.41--22.40&0.12--0.19&19.92--23.00&16.73--19.31&1.49--1.72&--&\textit{ab
initio}&\onlinecite{RubioPRB99}\\
Arroyo \textit{et al.}& 15.19& 0.40 &
18.08&10.85&0.69&--&FEM&\onlinecite{ArroyoPRB04}\\
Zhang \textit{et al.}&14.75&--&--&--&--&--&CTIP
&\onlinecite{ZhangIJSS02}\\
Caillerie \textit{et al.}&17.31&0.26&18.57&13.74&--&--&CTIP&\onlinecite{CaillerieJE06}\\
Wang & 34.63 or 17.31 &--&--&--& 1.12 or 1.21 &
--&CTIP&\onlinecite{WangIJSS04}\\
Present work &22.35&0.16&22.97&19.19&1.62&0.72&LDA&--
\end{tabular}
\end{ruledtabular}
\end{table*}

We should emphasize that our formula (\ref{freeng-gr}) holds
approximate up to the order of $(r_0/\rho)^2$ for SWNTs, where $r_0$
is the C-C length and $\rho$ the radius of the SWNT. The omitted
terms is in the order of $(r_0/\rho)^4$. This is the main reason for
the size effect on the elastic constants in the very small SWNTs
found in
Refs.~\onlinecite{ZhengQPRL05,ChangJMPS03,LiChouPRB04,ChangAPL05,MoriJJAP05}.
Additionally, we have not considered the effect of Stone-Wales
defects on the local properties of carbon nanotubes. In terms of
Refs.~\onlinecite{ChandraPRB04} and \onlinecite{BhattacharyaNT05},
we can deduce that the defects reduce the the elastic constants of
carbon nanotubes.

\subsubsection{Revisit the stability of SWNTs}
Now we will revisit the stability of SWNTs with the four fundamental
quantities $k_c$, $\bar{k}$, $k_d$, and $\tilde{k}$ or the
corresponding reduced quantities.

\begin{figure}[pth!]
\includegraphics[width=6cm]{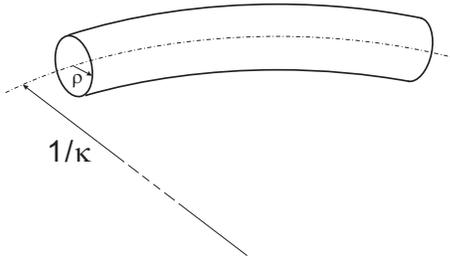}\caption{\label{figbendswnt}
Bent SWNT. $\rho$ and $1/\kappa$ are the radii of the SWNT and the
centerline of the SWNT, respectively.}
\end{figure}

First, let us consider a bent SWNT as shown in
Fig.~\ref{figbendswnt} where $\rho$ and $1/\kappa$ are the radii of
the SWNT and the centerline of the SWNT, respectively. Assume that
the centerline of the SWNT is not extended and the cross section of
the SWNT is still flat after bending under the condition $\rho\ll
L\ll 1/\kappa$, where $L$ is the total length of the centerline. In
terms of Eq.~(\ref{freeng-gr}), we can derive the deformation energy
due to bending as
\begin{equation}\Delta \mathcal{F} \approx \int_0^L (k_{rod}/2)\kappa^2 ds,\end{equation}
where $ds$ is the arc length element of the centerline. The bending
modulus of the rod is $k_{rod}=\pi \rho [( 2-\tilde{k}/k_{d})
\tilde{k}\rho ^{2}+k_{c}]$. For the SWNT with diameter in the order
of 1~nm, we can estimate $k_c\ll ( 2-\tilde{k}/k_{d}) \tilde{k}\rho
^{2}$. Considering Eq.~(\ref{Youngsmodip}), we have
\begin{equation}k_{rod}\approx\pi Y_s\rho^3.\end{equation}
If an axial compression force $F$ is loaded on the both ends of the
SWNT, following Euler rod theory,\cite{Love44} we can easily derive
the critical force, above which the SWNT is instable, as
\begin{equation}F_{c}^{rod}=2\alpha\pi k_{rod}/L^2,\end{equation}
where $\alpha$ depends on the boundary conditions in two ends of the
SWNT. Defining the critical strain as
$\varepsilon_{zc}^{rod}=F_{c}^{rod}/2\pi \rho Y_s$ and considering
the above two equations, we can derive
\begin{equation}\varepsilon_{zc}^{rod}= \alpha (\rho/L)^2.\label{epsceulerax2}\end{equation}
This relation has the same asymptotic behavior as
Eq.~(\ref{epsceulerax}), which, as mentioned above, has been
confirmed by a lot of theoretical and numerical researches.

Secondly, let us consider a short SWNT with radius $\rho$ and an
axial compression force loaded on its two ends. The force per length
along the circumference is denoted as $f$. Following Ru's work,
\cite{RuPRB2000} considering Eq.~(\ref{freeng-gr}) we have the
critical axial force density as
\begin{equation}f_c^{shell}=\alpha \sqrt{k_{c}Y_{s}}/\rho,\end{equation}
and the corresponding critical strain
\begin{equation}\varepsilon_{zc}^{shell}\equiv{f_{c}}/{Y_{s}}=(\alpha /\rho )\sqrt{k_{c}/Y_{s}},\label{epscshellax2}\end{equation}
where $\alpha$ depends on the boundary conditions in two ends of the
SWNT. $Y_s$ is the in-plane Young's modulus as shown in
Eq.~(\ref{Youngsmodip}). The above relation (\ref{epscshellax2}) has
the same asymptotic behavior as Eq.~(\ref{epscshellax}), which, as
mentioned in above, has been confirmed by several theoretical and
numerical researches.

Thirdly, let us consider a long enough SWNT with radius $\rho$ and a
radial compression pressure $p$ loaded on its surface. In terms of
the similar method on the stability of cell membranes, we can derive
the critical pressure
\begin{equation}p_{cr}^{shell}=3k_c/\rho^3,\label{pcrshellrd22}\end{equation}
above which the SWNT will lose its stability. This relation has the
same asymptotic behavior as Eq.~(\ref{pcrshellrd}). The
corresponding critical circumferential strain is
\begin{equation}\varepsilon_{c}^{cir}= 2\rho p_{cr}^{shell}/Y_s=6k_c/Y_s\rho^2.\label{epscrshellrd2}\end{equation}
Comparing Eq.~(\ref{pcrshellrd22}) with (\ref{criticalptube}), one
can find that the critical pressures for carbon nanotubes and lipid
tubules are in the same form. Yin \textit{et al.} have noticed this
similarity in the recent work \onlinecite{YinIEE2006}. However, the
profound mechanism is different: nanotubes can endure the shear
strain while lipid tubules cannot.

It seems that no literature discusses the possible instability of a
SWNT under axial tension. Here we will give a qualitative analysis.
Assume the tension density (i.e., force per length) to be $f$. The
axial strain under the tension is $f/Y_s$ and the corresponding
circumferential strain is $\nu_s f/Y_s$. When it is beyond the
critical value (\ref{epscrshellrd2}), the SWNT will be instable.
Thus we obtain the critical tension density
\begin{equation}f_c^{tsn}=6k_c/\nu_s\rho^2.\end{equation}
Only if $f_c^{tsn}$ is below the strength of the SWNT, the buckling
phenomenon under tension can be observed.

Till now, we have not found that $\bar{k}$ exists explicitly in the
above equations (\ref{epsceulerax2})--(\ref{epscrshellrd2}) for
nanotubes. Because the term related to $\bar{k}$ in the free energy
(\ref{freeng-gr}) can be transformed into the boundary term with the
aid of Gauss-Bonnet formula, $\bar{k}$ should be implicitly
contained by $\alpha$ in these equations, which need the further
investigations in the future.

\section{Conclusion and prospect\label{SecConclusion}}
In summary, we present the elastic theory of low-dimensional (one-
and two-dimensional) continua and its applications in bio- and
nano-structures. The elastic theory of Kirchhoff rod, Helfrich rod,
bending-soften rod, fluid membrane, and solid shell is revisited. We
construct the free energy density of the continua on the basis of
the symmetry argument. The fundamental equations can be derived from
the bottom-up and the top-down standpoints. Although they have
different forms obtained from these two viewpoints, several examples
reveal that they are, in fact, equivalent to each other. We
investigate the kink stability of short DNA rings, the elasticity of
lipid membranes, and the adhesions between a vesicle and a substrate
or another vesicle. A cell membrane is simplified as a composite
shell of lipid bilayer and membrane skeleton. The membrane skeleton
is shown to enhance highly the mechanical stability of cell
membranes. We propose a revised Lenosky lattice model based on the
local density approximation and derive its continuum form up to the
second order terms of curvatures and strains, which is the same as
the free energy of 2D solid shells. The intrinsic roughening of
graphene and several typical mechanical properties of carbon
nanotubes are addressed by using this continuum form. We can abandon
the controversial thickness and Young's modulus of graphene and
SWNTs if we adopt this continuum form to describe the mechanical
properties of graphene and SWNTs.

Finally, we would like to list a few open problems which need to be
addressed in the future work.

(i) The vesicles with lipid domains have been investigated in
Sec.~\ref{Seclipm}. There is a special lipid domain at
liquid-ordered phase, so called the raft, which is enriched in
cholesterol and sphingolipids. Cholesterol is a kind of chiral lipid
molecules, which has not been included in the previous and present
theory of lipid domains. A new theory with the chirality on the raft
domain should be developed.

(ii) The composite shell, as a model of cell membranes, has been
investigated in Sec.~\ref{Seccellmb} where the constraint between
the area of the lipid bilayer and membrane skeleton is totally
neglected. Additionally, only the small deformation of cell
membranes are addressed in this review. The large deformation
behavior of cell membranes \cite{BoeyBJ98,DischerBJ98,LimPNAS2002}
has recently discussed through numerical simulations. It is
necessary to reconsider the composite shell model with the
constraint $\int J dA=\int J_b dA$ and its behavior under large
deformation theoretically.

(iii) We suggest adopting four parameters $k_c$, $\bar{k}$, $k_d$,
and $\tilde{k}$ to describe the mechanics of graphitic structures in
Sec.~\ref{SecApNanoStuct}. However, there are sparse studies on
$\bar{k}$ in previous literature. It is highly expected to
theoretical and experimental work on this quantity.

(iv) We only talk about the thermal fluctuation on the discussion of
graphene. The fluctuations of DNA, lipid membranes, and cell
membranes are not in the range of our topics, on which we suggest
that gentle readers consult
Refs.\onlinecite{PanyukovPRE2001,Seifert97,MarkoSiggia95,HaijunPRL99}.

(v) The elastic theory presented in this review is a static theory.
Thus we are very regretted that we have to omit several important
subjects such as the vesicles in shear
flows,\cite{KrausPRL96,FinkenJPC06,Noguchipnas05,MisbahPRL06,SkotheimPRL07}
and dynamic response of carbon nanotubes or nanotube networks
\cite{WangDaiC06,ColuciPRB07} and so on. These topics will be
quickly developed in the future.

\section*{Acknowledgements}
We are very grateful to Dr. Q. X. Li and Prof. T. Lenosky for their
help in our DFT calculations. We thank Prof. G.-L. Xu because he let
us know the work by Giaquinta and Hildebrandt, which we have not
noticed before. We are grateful to Prof. Z.-C. Zhou for his kind
comments. Some materials in this review are prepared in Tamkang
University where ZCT is supported by the National Science Council
(grant no. NSC 94-2119-M-032-010), and the others are prepared in
Universit\"{a}t Stuttgart where ZCT is supported by the Alexander
von Humboldt foundation. ZCT is also grateful to the support of
Nature Science Foundation of China (grant no. 10704009).

\section*{\textit{Note added in proof}} After this review was in press, we
noticed that recent researches$^{248,249}$ on the mechanical
properties of nanosprings$^{248}$ and amorphous straight
nanowires$^{249}$ within the framework of Kirchhoff rod. We were
also informed of the researches$^{250,251}$ by Arroyo and Belytschko
on discussing the buckling pattern of multi-walled carbon nanotubes
under pure bending. Additionally, we emphasize that the similar
equations to (114) and (115) without $p$ first obtained by Zhang et
al.$^{252}$

\appendix
\section{Curve variational theory\label{SecAppvarcc}}
Here we sketch merely the derivation of Eqs.~(\ref{ELKrod2}) and
(\ref{ELKrod3}) because the derivation of Eq.~(\ref{ELKrod1}) is
trivial.

For convenience, we denote $\mathbf{e}_{1}=\mathbf{N}$,
$\mathbf{e}_{2}=\mathbf{B}$, $\mathbf{e}_{3}=\mathbf{T}$. Then we
have $d\mathbf{r}=\omega_3\mathbf{e}_{3}$ and
$d\mathbf{e}_i=\omega_{ij}\mathbf{e}_j$ where
\begin{equation}\omega_3=ds,\omega _{12}=\tau \omega _{3},\omega
_{13}=-\kappa \omega _{3},\omega
_{23}=0.\label{appvarc0}\end{equation} Following the spirit of
Ref.~\onlinecite{TuJPA04}, any infinitesimal deformation of a curve
can be achieved by a displacement vector at
each point on the curve as \begin{equation}\delta \mathbf{r}\equiv \mathbf{v}=\Omega_{1}\mathbf{e}_{1}+\Omega_{2}%
\mathbf{e}_{2}+\Omega_{3}\mathbf{e}_{3},\end{equation} where
$\delta$ can be understood a variational operator. The frame is also
changed because of the deformation of the curve, which is denoted as
\begin{equation}\delta\mathbf{e}_i=\Omega_{ij}\mathbf{e}_j,\quad (i=1,2,3),\label{varyei}\end{equation}
where $\Omega_{ij}=-\Omega_{ji},(i,j=1,2,3)$ corresponds to the
rotation of the frame due to the deformation of the curve. From
$\delta d\mathbf{r}=d\delta\mathbf{r}$, $\delta
d\mathbf{e}_j=d\delta\mathbf{e}_j$, and $\delta
d\phi=d\delta\phi=0$, we can derive
\begin{eqnarray}
&&\delta \omega _{3} =\Omega _{1}\omega _{13}+\Omega _{2}\omega
_{23}+d\Omega _{3},\label{appvarc1} \\
&&\omega _{3}\Omega _{31} =d\Omega _{1}+\Omega _{2}\omega
_{21}+\Omega
_{3}\omega _{31}, \\
&&\omega _{3}\Omega _{32} =\Omega _{1}\omega _{12}+d\Omega
_{2}+\Omega
_{3}\omega _{32}, \\
&&\delta \omega _{ij} =d\Omega _{ij}+\Omega _{il}\omega _{lj}-\omega
_{il}\Omega
_{lj},\\
&&\delta\phi'\,ds=-\phi'\delta\omega_3.\label{appvarc5}
\end{eqnarray}
Considering the above equations (\ref{appvarc0}),
(\ref{appvarc1})--(\ref{appvarc5}) and integral by parts as well as
Stokes' theorem, we can derive Eqs.~(\ref{ELKrod2}) and
(\ref{ELKrod3}) from the free energy (\ref{freeng-rod}).

\section{Surface variational theory \label{varisurf}}
Following the spirit of Ref.~\onlinecite{TuJPA04}, any infinitesimal
deformation of a surface can be achieved by a displacement vector at
each point on the surface as \begin{equation}\delta \mathbf{r}\equiv \mathbf{v}=\Omega_{1}\mathbf{e}_{1}+\Omega_{2}%
\mathbf{e}_{2}+\Omega_{3}\mathbf{e}_{3},\end{equation} where
$\delta$ can be understood a variational operator. The frame is also
changed because of the deformation of the surface, which is still
denoted as
\begin{equation}\delta\mathbf{e}_i=\Omega_{ij}\mathbf{e}_j,\quad (i=1,2,3),\label{varyei}\end{equation}
where $\Omega_{ij}=-\Omega_{ji},(i,j=1,2,3)$ corresponds to the
rotation of the frame due to the deformation of the surface. From
$\delta d\mathbf{r}=d\delta\mathbf{r}$, $\delta
d\mathbf{e}_j=d\delta\mathbf{e}_j$, and
Eqs.~(\ref{sframer})--(\ref{omega13}), we can derive
\begin{eqnarray}\delta \omega _{1}&=&d\mathbf{v}\cdot \mathbf{e}_{1}-\omega _{2}\Omega _{21},\label{surfacev1}\\
\delta \omega _{2} &=&d\mathbf{v}\cdot \mathbf{e}_{2}-\omega
_{1}\Omega _{12},\\
\Omega_{13} &=&\Omega_{3,1}+a\Omega_{1}+b\Omega_{2},\\
\Omega_{23} &=&\Omega_{3,2}+b\Omega_{1}+c\Omega_{2},\\
\delta \omega _{ij}&=&d\Omega _{ij}+\Omega _{il}\omega _{lj}-\omega
_{il}\Omega _{lj}.\label{surfacev4}\end{eqnarray} These equations
are the essential equations of the surface variational theory based
on the moving frame method. With them as well as
Eqs.~(\ref{omega13}) and (\ref{relat2HK}), we can easily derive
\begin{eqnarray}&&\delta dA =( \mathrm{div\,}\mathbf{v}-2H\Omega _{3}) dA,\\
&&\delta (2H) =[\nabla ^{2}+(4H^{2}-2K)] \Omega_{3}+\nabla (2H)
\cdot \mathbf{v},\\
&&\delta K =\nabla \cdot \tilde{\nabla}\Omega _{3}+2KH\Omega
_{3}+\nabla K\cdot \mathbf{v}.
\end{eqnarray} Using the above three equations and the Stokes'
theorem, we can easily derive Eq.~(\ref{ELclosFM}) from the free
energy (\ref{frengFM}), or
Eqs.~(\ref{shapeopenLV})--(\ref{openbound3}) from the free energy
(\ref{FrengopenV}), and so on.

\section{Stokes' theorem and the other important geometric relations}
The Stokes' theorem is a crucial theorem in differential geometry.
Let us denote the boundary of domain $\mathfrak{D}$ as
$\partial\mathfrak{D}$. The Stokes' theorem states as: If $\omega$
is a differential form on $\partial\mathfrak{D}$, then
\begin{equation}\oint_{\partial\mathfrak{D}}\omega=\int_\mathfrak{D} d\omega.\end{equation}
In particular, $\int_\mathfrak{D} d\omega=0$ for a closed domain
$\mathfrak{D}$.

It contains a lot of geometric relations, which are listed as
follows.

(i) For any smooth functions $f$ and $h$ on 2D domain
$\mathfrak{D}$, we have
\begin{eqnarray}\int_\mathfrak{D}(fd{\ast}dh-hd{\ast}df)=\oint_{\partial
\mathfrak{D}}(f{\ast} dh-h{\ast}
df),\\
\int_\mathfrak{D}(fd{\ast}\tilde{d}h-hd{\ast}\tilde{d}f)=\oint_{\partial
\mathfrak{D}}(f{\ast} \tilde{d}h-h{\ast}
\tilde{d}f),\\
\int_\mathfrak{D}(fd\tilde{\ast}\tilde{d}h-hd\tilde{\ast}\tilde{d}f)=\oint_{\partial
\mathfrak{D}}(f\tilde{\ast} \tilde{d}h-h\tilde{\ast}
\tilde{d}f).\end{eqnarray} where $\tilde{d}$ and $\tilde{\ast}$ are
generalized differential operator and Hodge star which satisfy
$\tilde{d}f= f_1\omega_{13}+f_2\omega_{23}$ and
$\tilde{\ast}\tilde{d}f= f_1\omega_{23}-f_2\omega_{13}$ if
$df=f_1\omega_{1}+f_2\omega_{2}$.\cite{TuJPA04}

(ii) If $\mathbf{u}$ is a vector defined on a closed surface, then
\begin{equation}\int d\mathbf{u}\cdot\wedge \ast d\mathbf{u}=-\int
\mathbf{u}\cdot d\ast d\mathbf{u},\end{equation} where the dot
represents the inner product of vectors.

(iii) For the tensors $\mathfrak{S}$ and $\mathfrak{M}$ defined in
Sec.~\ref{ETLDC}, we have
\begin{eqnarray}\oint_{\partial\mathfrak{D}}\mathfrak{S}\cdot \mathbf{b}\, ds=\int_{\mathfrak{D}}\mathrm{div\,} \mathfrak{S}\,dA,\\
\oint_{\partial\mathfrak{D}}\mathfrak{M}\cdot \mathbf{b}\,
ds=\int_{\mathfrak{D}}\mathrm{div\,} \mathfrak{M}\,dA,\end{eqnarray}
where $\mathfrak{D}$ is a 2D domain with boundary
$\partial\mathfrak{D}$ and $\mathbf{b}$ is the normal vector of
$\partial\mathfrak{D}$ in the tangent plane.

The above three items are also called the Stokes' theorem in this
review, which are widely used in the variational process.

The other geometric identities linking the vector form and
differential form on a smooth surface used in this review are
summarized as follows without additional proof.
\begin{eqnarray}
&&(\mathrm{curl\,} \mathbf{u})\,dA=d(\mathbf{u}\cdot d\mathbf{r}),\\
&&(\mathrm{div\,} \mathbf{u})\,dA=d(\ast\mathbf{u}\cdot d\mathbf{r}),\\
&&(\tilde{\nabla} \cdot \mathbf{u})\,dA=d(\tilde{\ast}\mathbf{u}\cdot \tilde{d}\mathbf{r}),\\
&&(\bar{\nabla} \cdot \mathbf{u})\,dA=d(\ast\mathbf{u}\cdot \tilde{d}\mathbf{r}),\\
&&\nabla f \cdot d\mathbf{r}=df,\\
&&\tilde{\nabla} f \cdot d\mathbf{r}=\tilde{d}f,\\
&&(\nabla^2f)\,dA=d\ast df,\\
&&(\nabla\cdot\bar{\nabla} f)\,dA=d\ast\tilde{d}f,\\
&&(\nabla\cdot\tilde{\nabla} f)\,dA=d\tilde{\ast} \tilde{d}f,\\
&&(\nabla f \cdot\mathbf{u})\,dA= df\wedge\ast\mathbf{u}\cdot
d\mathbf{r},\\
&&(\nabla^2\mathbf{u})\,dA=d\ast d\mathbf{u},\\
&&(\nabla\mathbf{u})\cdot\,d\mathbf{r}=d\mathbf{u}.
\end{eqnarray}


\begin{thebibliography}{}
\bibitem{remktidk}$\tilde{k}$ in this review is twice of the tradtional convention
of the shear modulus.
\end{thebibliography}
\end{document}